\documentclass[prd,twocolumn,superscriptaddress,floatfix,nofootinbib]{revtex4-2}
\pdfoutput=1 
\usepackage[T1]{fontenc} 
\usepackage{amsmath,amssymb,amsfonts,amsthm}
\usepackage{graphicx}
\usepackage[usenames,dvipsnames]{color}
\usepackage{enumitem}
\usepackage{verbatim}
\usepackage[percent]{overpic}
\usepackage{rotating}
\usepackage{hyperref}
\usepackage{array}
\usepackage{mathtools}
\usepackage{lmodern}
\usepackage[normalem]{ulem}
\usepackage{braket}
\usepackage{tensor}
\usepackage{dsfont}
\usepackage{bm}
\usepackage{tikz}

\usepackage{mathrsfs}

\usetikzlibrary{decorations.pathmorphing}
\usepackage{CJKutf8}

\newcommand{\kgy}[1]{{\color{blue}\bf [Ken: #1]}}

\usetikzlibrary{positioning}
\usetikzlibrary{calc,through,backgrounds}

\theoremstyle{definition}
\newtheorem{definition}{Definition}[section]
\newtheorem{theorem}{Theorem}[section]

\newcommand{\kako}[1]{\left( #1 \right)}
\newcommand{\kagikako}[1]{\left[ #1 \right]}
\newcommand{\ts}[1]{ _{\text{#1}} }

\newcommand{\Bigkagikako}[1]{\Big[ #1 \Big]}

\allowdisplaybreaks

\DeclareMathOperator{\Tr}{Tr}

\newcommand{\dd}{\text{d}}

\newcommand{\bx}{{\bm{x}}}

\newcommand{\id}{\mathds{1}}
\newcommand{\sx}{\mathsf{x}}

\newcommand{\ii}{\mathsf{i}}

\newcommand{\ads}[1]{\text{AdS}_{#1}}

\begin{document}

\title{Tripartite Entanglement Extraction from the Black Hole Vacuum}

\author{Ireneo James Membrere}
\email{imembrere@uwaterloo.ca} 
\affiliation{Department of Physics and Astronomy, University of Waterloo, Waterloo, Ontario, N2L 3G1, Canada}

\author{Kensuke Gallock-Yoshimura}
\email{kgallock@uwaterloo.ca} 

\affiliation{Department of Physics and Astronomy, University of Waterloo, Waterloo, Ontario, N2L 3G1, Canada}


\author{Laura J. Henderson}
\email[]{laura.henderson@uq.edu.au}
\affiliation{Centre for Engineered Quantum Systems, School of Mathematics and Physics,
The University of Queensland, St. Lucia, Queensland, 4072, Australia}
\affiliation{Department of Physics and Astronomy, University of Waterloo, Waterloo, Ontario, N2L 3G1, Canada}

\author{Robert B. Mann}
\email{rbmann@uwaterloo.ca}
\affiliation{Department of Physics and Astronomy, University of Waterloo, Waterloo, Ontario, N2L 3G1, Canada}

\begin{abstract}
The first investigation of tripartite entanglement harvesting in the vicinity of a black hole is carried out. 
Working in the context of a static Ba\~{n}ados-Teitelboim-Zanelli (BTZ) black hole spacetime we find that it is possible to harvest tripartite entanglement in regions where harvesting of bipartite entanglement is known to be impossible due to intense Hawking radiation. 
In these situations, it implies that the harvested entanglement is of the Greenberger-Horne-Zeilinger (GHZ) type. 
\end{abstract}

\maketitle
\flushbottom

\section{Introduction}

Entanglement in quantum fields has been a subject of considerable interest in recent years. 
A varied range of sources motivate this, including the origin of black hole entropy \cite{Bombelli:1986rw,Srednicki:1993im}, nonclassical states of light \cite{RevModPhys.84.621}, critical phenomena in condensed matter systems \cite{Osterloh_2002,RevModPhys.80.517}, the anti-de Sitter/conformal field theory correspondence \cite{PhysRevLett.96.181602}, and the black hole information paradox \cite{Raju.lessons.BHinfo}.

A quantum field at two points in spacetime is known to be entangled \cite{summers1985bell, summers1987bell}, and this entanglement depends on the background curved spacetime. 
Operationally, it can be probed using the so-called \textit{entanglement harvesting protocol}. 
The protocol utilizes multiple qubits that locally interact with a quantum field; after some time they extract correlations from the field and become entangled with each other,  even if they are causally disconnected \cite{Valentini1991nonlocalcorr, reznik2003entanglement, reznik2005violating, pozas2015harvesting, TjoaSignal}. 
Such qubits are often modeled by a nonrelativistic first-quantized two-level system known as the Unruh-DeWitt (UDW) detector model \cite{Unruh1979evaporation, DeWitt1979}, which is known to represent essential features of the light-matter interaction so long as angular momentum is not exchanged \cite{EMM.wavepacket, Alhambra.CasimirForces, pozas2016entanglement}.  
The extracted entanglement depends not only on the state of motion of the detectors but also on the background spacetime, thereby providing a probe of how gravity affects the quantum vacuum \cite{smith2016topology, kukita2017harvesting, henderson2018harvesting, ng2018AdS, henderson2019entangling, cong2020horizon, FinnShockwave}.

Among these investigations, entanglement harvesting in black hole spacetimes exhibits particularly interesting phenomena. 
Consider two UDW detectors in a black hole spacetime, hovering at a fixed radius from an event horizon. 
It was established some time ago  \cite{henderson2018harvesting} that if one of the detectors is close enough to the event horizon, then bipartite entanglement cannot be harvested due to a combination of gravitational redshift suppressing nonlocal correlations and increased Hawking radiation amplifying detector noise. 
The region that inhibits extraction of entanglement is called the \textit{entanglement shadow}, and has subsequently been seen in various kinds of black hole spacetimes \cite{robbins2020entanglement, Tjoa2020vaidya, Ken.Freefall.PhysRevD.104.025001,Henderson:2022oyd,Kendra.BTZ,Barman:2023rhd}.

Comparatively little is known about tripartite entanglement harvesting. 
This phenomenon was first examined by Silman and Reznik \cite{Silman.Wstate}, who showed that detectors can extract tripartite entanglement that can be  distilled into a W state. 
Some time later an investigation of tripartite harvesting in (1+1) dimensions using Gaussian quantum mechanics was carried out \cite{Lorek2014tripartite}, and recently a study in flat spacetime \cite{DianaGaussianTripartite} found that no bipartite entanglement was needed to harvest tripartite entanglement; furthermore, even if one detector is at  a significant distance from the other two, tripartite entanglement can be harvested. 
Another recent study~\cite{DianaSingleDeltaTripartite} showed that three detectors instantaneously interacting with a field can obtain Greenberger-Horne-Zeilinger (GHZ)-type entanglement, despite the fact that bipartite entanglement harvesting in this scenario is prohibited by a no-go theorem \cite{pozas2017degenerate, Simidzija.Nonperturbative, Simidzija2018no-go}.

Motivated by the above, we ask the following question: what will happen to the entanglement shadow of a black hole if another detector is introduced? 
We are particularly interested if and where detectors can harvest tripartite entanglement even though bipartite harvesting is suppressed (or even forbidden). 
For this reason entanglement harvesting in a black hole spacetime is a suitable setting because of the existence of the bipartite entanglement shadow.

We consider three UDW detectors placed in various static configurations outside a  Ba\~{n}ados-Teitelboim-Zanelli (BTZ) black hole. 
This choice of black hole is in large part motivated by simplicity: the Wightman function can be expressed as an image sum of the correlation function in anti-de Sitter (AdS)  spacetime \cite{LifschytzBTZ, carlipBTZ}. 
It was also the original setting in which the entanglement shadow was discovered \cite{henderson2018harvesting} and so makes a comparison to the bipartite case straightforward.

We shall assume that the detectors interact with a quantum scalar field governed by a Gaussian switching function. 
We find that black holes affect the tripartite properties of the quantum vacuum in a manner quite distinct from the bipartite case. 
Specifically, we indeed find that tripartite entanglement harvesting is possible in regions where bipartite harvesting is not. 
This in turn implies that the harvested entanglement is of the GHZ type \footnote{We note that GHZ-entangled states have recently been shown to play a useful role in understanding the black hole information paradox \cite{e22121387}}.

The outline of our paper is as follows. 
In section \ref{sec2}, we review the basic properties of the BTZ black hole and the UDW detector model in the tripartite case. 
We also outline why the $\pi$-tangle is an appropriate measure of tripartite entanglement for our investigation. 
In section \ref{sec3} we present our results, providing a comparison with the bipartite case as appropriate. 
We summarize our results in a concluding section.

\section{Particle detectors}
\label{sec2}

\subsection{Quantum fields in BTZ spacetime}
The $(2+1)$-dimensional BTZ black hole spacetime \cite{BTZ1, BTZ2} can be obtained as a solution to the Einstein field equations with a negative cosmological constant $\Lambda=-1/\ell^2$, where $\ell (\geq 0)$ is the AdS length. 
The line-element reads
\begin{align}
\label{btzmet}
    \dd s^2
    &=
        - f(r) \dd t^2
        + \dfrac{ \dd r^2 }{f(r)}
        + r^2 
        \dd \varphi^2\,,
\end{align} 
where $f(r)=r^2/\ell^2  -M$, $t\in \mathbb{R}, r\in (0,\infty)$, and $\varphi \in [0, 2\pi)$. 
Here, $M$ is the (dimensionless) mass of the black hole. 
The event horizon is located at $r=r_h \equiv \ell \sqrt{M}$.

We note that the BTZ spacetime can be obtained from $\ads{3}$ spacetime by a topological identification and thereby BTZ spacetime is conformally equivalent to $\ads{3}$. 
This fact will be important when one considers a quantum field in BTZ spacetime, in which the correlation functions can be written as a sum of correlators in $\ads{3}$. 
It is this feature that makes the BTZ black hole particularly attractive for studying the relationship between gravitation and entanglement in quantum fields.

Let us now consider a quantum scalar field $\hat \phi(\sx)$ defined on this spacetime. 
Specifically, we consider a massless conformally coupled scalar field satisfying the Klein-Gordon equation, 
\begin{align}
    (\square - R/8) \hat \phi(\sx)=0\,,
\end{align}
where $\square$ is the d'Alembert operator and $R$ is the Ricci scalar. 
Since BTZ spacetime is conformally equivalent to $\ads{3}$, we are able to consider field correlations in terms of the $\ads{3}$ 
 Wightman function  $W(\sx, \sx')\coloneqq \braket{0|\hat \phi(\sx) \hat \phi(\sx')|0}$,  
which is the two-point correlation function 
in a vacuum state $\ket{0}$ of the field.

In what follows, we will always consider the Hartle-Hawking vacuum so that the BTZ black hole is in thermal equilibrium with its exterior. 
Then the Wightman function in BTZ spacetime, $W\ts{BTZ}(\sx, \sx')$, can be expressed as an image sum of the correlation function in $\ads{3}$ spacetime \cite{LifschytzBTZ, carlipBTZ}: 
\begin{align}
    &W\ts{BTZ}(\sx, \sx')
    =
        \sum_{n=-\infty}^\infty
        W_{\ads{3}}(\sx, \Gamma^n \sx') \notag \\
    &=
        \dfrac{1}{ 4\pi \sqrt{2}\ell }
        \sum_{n=-\infty}^\infty
        \kagikako{
            \dfrac{1}{ \sqrt{ \sigma_\epsilon (\sx, \Gamma^n \sx') } }
            -
            \dfrac{\zeta}{ \sqrt{ \sigma_\epsilon (\sx, \Gamma^n \sx')+2 } }
        }, \label{eq:BTZ Wightman}
\end{align}
where $\Gamma:(t,r,\varphi)\mapsto (t,r,\varphi+ 2\pi)$ is the topological identification that changes the $\ads{3}$ metric \eqref{btzmet} from that of AdS-Rindler space [in which $\varphi$ is not identified with any period, so that $\varphi\in (-\infty,\infty)$] into a BTZ black hole. 
The quantity
\begin{align}
    \sigma_\epsilon (\sx, \Gamma^n \sx')
    &= 
        \dfrac{r r' }{ r_h^2} 
        \cosh 
        \kagikako{
            \dfrac{r_h}{\ell} (\Delta \varphi - 2\pi n )
        }
        -1 \label{sigma} \\
        &-\dfrac{ \sqrt{ (r^2-r_h^2) (r^{\prime 2}-r_h^2) } }{ r_h^2 } 
        \cosh 
        \kako{
            \dfrac{r_h}{\ell^2} 
            \Delta t - \ii \epsilon
\nonumber        }
\end{align}
is the geodesic separation between the two points,
with $\Delta \varphi\coloneqq \varphi - \varphi', \Delta t\coloneqq t-t'$, and $\epsilon$ is a UV cutoff. 
Since the spatial infinity of BTZ spacetime is timelike, one needs to impose a boundary condition, which is characterized by $\zeta$. 
Throughout the paper, we choose the Dirichlet boundary condition, $\zeta=1$.

In what follows we will find it useful to  introduce a proper distance 
\begin{align}
    d(r_1, r_2)
    &=
        \ell \ln 
        \kako{
            \dfrac{ r_2 + \sqrt{ r_2^2 - r_h^2 } }{ r_1 + \sqrt{ r_1^2 - r_h^2 } }
        }
\end{align}
on a time-slice $\dd t=0$ between two spacetime points $(t,r_1,\varphi)$ and $(t, r_2, \varphi)$ with $r_2 > r_1 \geq r_h$. 
We shall consider all detector separations in terms of this proper distance.


\subsection{UDW detector model}

A UDW detector is a first-quantized two-level system with an energy gap $\Omega$ between the ground $\ket{g}$ and excited $\ket{e}$ states. 
Consider three UDW detectors, labeled by $j \in \{ A,B,C \}$. 
Each of them shall  be taken to be pointlike and to locally interact with the field through a time-dependent coupling, $\lambda_j \chi_j(\tau_j/\sigma)$, where $\lambda_j$ is the coupling constant and $\chi_j(\tau_j/\sigma)$ is a switching function, with a characteristic width, $\sigma$, which is defined in terms of the proper time $\tau_j$ of detector-$j$.

The total interaction Hamiltonian generating a time-translation with respect to the common time $t$ is given by 
\begin{align}
    \hat H\ts{I}^t(t)
    &=
        \sum_{j \in \{A,B,C\}}
        \dfrac{\dd \tau_j}{ \dd t } \hat H^{\tau_j}_j(\tau_j(t))\,,
\end{align}
where 
\begin{align}
    \hat H^{\tau_j}_j(\tau_j)
    &\coloneqq
        \lambda_j \chi_j(\tau_j/\sigma) \hat \mu_j(\tau_j)
        \otimes 
        \hat \phi(\bx_j(\tau_j))
\end{align}
is an interaction Hamiltonian that describes the interaction between detector-$j$ and the field $\hat \phi$ along the trajectory of the detector. 
The superscript of a Hamiltonian indicates the time variable of the generated time-translation. 
The operator, $\hat \mu_j(\tau_j)$, is the monopole moment of detector-$j$ given by
\begin{align}
    \hat \mu_j(\tau_j) 
    &=
        \ket{e_j} \bra{g_j} e^{ \ii \Omega_j \tau_j }
        +
        \ket{g_j} \bra{e_j} e^{ -\ii \Omega_j \tau_j }\,.
\end{align}

Let $\mathcal{T}_t$ be a time-ordering symbol with respect to $t$. 
Assuming the  dimensionless coupling constant,  {$\tilde{\lambda}\coloneqq\lambda\sqrt{\sigma}$,} is small, we can perform the  Dyson series expansion  
\begin{align}
    \hat U\ts{I}
    &=
        \mathcal{T}_t 
        \exp 
        \kako{
            -\ii \int_{\mathbb{R}} \dd t\,\hat H\ts{I}^t(t)
        } \nonumber \\
    & =
        \id + \hat U^{(1)} + \hat U^{(2)} + \mathcal{O}(\lambda^3)\,,
\end{align}
where 
\begin{subequations}
\begin{align}
    \hat U^{(1)}
    &=
        -\ii \int_{-\infty}^\infty \dd t\,\hat H\ts{I}^t(t)\,,\\
    \hat U^{(2)}
    &=
        - \int_{-\infty}^\infty \dd t_1
        \int_{-\infty}^{t_1} \dd t_2\,
        \hat H\ts{I}^t(t_1) \hat H\ts{I}^t(t_2)\,.
\end{align}
\end{subequations}

We then assume that the detectors are all initially in their ground states and the field is in the Hartle-Hawking vacuum $\ket{0}$. 
The initial density operator is then 
\begin{align}
    \rho_0
    &=
        \ket{g}_A\bra{g} 
        \otimes 
        \ket{g}_B\bra{g}
        \otimes 
        \ket{g}_C\bra{g}
        \otimes 
        \ket{0}\bra{0}\,,
\end{align}
and the final density operator is 
\begin{align}
    \rho\ts{tot}
    &=
        \hat U\ts{I} \rho_0 \hat U\ts{I}^\dag\,.
\end{align}
The final density matrix for the three detectors, $\rho_{ABC}$, can be obtained by tracing out the field degree of freedom, 
\begin{align}
    \rho_{ABC}
    &=
        \Tr_\phi[ \rho\ts{tot} ]\,,
\end{align}
and it can be shown that this density matrix in the basis $\{ \ket{g_A g_B g_C}$, $\ket{g_A g_B e_C}$, $\ket{g_A e_B g_C}$, $\ket{e_A g_B g_C}$, $\ket{g_A e_B e_C}$, $\ket{e_A g_B e_C}, \ket{e_A e_B g_C}, \ket{e_A e_B e_C} \}$ has the form \cite{DianaGaussianTripartite}
\begin{align}
\rho_{ABC}
    &=
        \left[
        \begin{array}{cccccccc}
        r_{11} &0 &0 &0 &r_{51}^* &r_{61}^* &r_{71}^* &0  \\
        0 &r_{22} &r_{32}^* &r_{42}^* &0 &0 &0 &r_{82}^*  \\
        0 &r_{32} &r_{33} &r_{43}^* &0 &0 &0 &r_{83}^*  \\
        0 &r_{42} &r_{43} &r_{44} &0 &0 &0 &r_{84}^*  \\
        r_{51} &0 &0 &0 &r_{55} &r_{65}^* &r_{75}^* &0  \\
        r_{61} &0 &0 &0 &r_{65} &r_{66} &r_{76}^* &0  \\
        r_{71} &0 &0 &0 &r_{75} &r_{76} &r_{77} &0  \\
        0 &r_{82} &r_{83} &r_{84} &0 &0 &0 &r_{88} 
        \end{array}
        \right]\,. \label{eq:rhoABC}
\end{align}
For our perturbative analysis, the density matrix  reads 
\begin{widetext}
\begin{align}
    \rho_{ABC} 
    = 
        \left[
        \begin{array}{cccccccc}
		1-(P_A+P_B+P_C) & 0 & 0 & 0 & X_{BC}^*  & X_{AC}^* & X_{AB}^* & 0 \\
		0 & P_C & C_{BC}^*  & C_{AC}^* & 0 & 0 & 0 & 0 \\
		0 & C_{BC} & P_B & C_{AB}^*  & 0  & 0 & 0 & 0 \\
		0 & C_{AC} & C_{AB}  & P_A  & 0 & 0 & 0 & 0 \\
		X_{BC} & 0 & 0 & 0 & 0 & 0 & 0 & 0 \\
		X_{AC} & 0 & 0 & 0 & 0 & 0 & 0 & 0 \\ 
		X_{AB} & 0 & 0 & 0 & 0 & 0 & 0 & 0 \\
		0 & 0 & 0 & 0 & 0 & 0 & 0 & 0
	    \end{array}
        \right]
	    + \mathcal{O}(\lambda^4)\,, \label{eq:rhoABC2ndOrder}
\end{align}
to  leading order in $\lambda$, where 
\begin{subequations}
\begin{align}
    P_j 
    &= 
        \lambda_j^2 
        \int_{\mathbb{R}} \dd\tau_j 
        \int_{\mathbb{R}} \dd \tau_j'\, 
        \chi_j(\tau_j/\sigma) \chi_j(\tau_j'/\sigma) 
        e^{-\ii \Omega_j(\tau_j -\tau_j')} 
        W(x_j(\tau_j),x_j(\tau_j'))\,,\label{eqn:DetectorProbability} \\
    C_{jk}
    &= 
        \lambda_j \lambda_k
        \int_{\mathbb{R}} \dd \tau_j 
        \int_{\mathbb{R}} \dd \tau_k'\, 
        \chi_j(\tau_j/\sigma) \chi_{k}(\tau_{k}'/\sigma) 
        e^{-\ii(\Omega_j \tau_j - \Omega_{k} \tau_{k}')} W(x_j(\tau_j),x_{k}(\tau_{k}'))\,, \\
    X_{jk}
    &= 
        -\lambda_j \lambda_k 
        \int_{\mathbb{R}} \dd\tau_j 
        \int_{\mathbb{R}} \dd \tau_{k}'\, 
        \chi_j(\tau_j/\sigma) \chi_{k}(\tau_{k}'/\sigma) 
        e^{\ii (\Omega_j \tau_j + \Omega_{k} \tau_{k}')} \notag \\
        &\quad\times 
        \Bigkagikako{
            \theta(t(\tau_j)-t(\tau_{k}'))
            W(x_j(\tau_j),x_{k}(\tau_{k}'))
            +
            \theta(t(\tau_{k}')-t(\tau_{j}))
            W(x_{k}(\tau_{k}'), x_j(\tau_j))
        }
\end{align}\label{eq:elements general}
\end{subequations}
\end{widetext}
respectively correspond to detector response, pairwise detector correlations, and pairwise nonlocal correlations. 
Here $\theta(t)$ is the Heaviside step function, which depends on $t(\tau)$, and $W=W\ts{BTZ}$.

The reduced density matrix $\rho_{jk}$ for detectors-$j$ and $k$ reads
\begin{align}
    \rho_{jk}
    &=
        \left[
        \begin{array}{cccc}
        1-P_j-P_k &0 &0 &X_{jk}^*  \\
        0 &P_k &C_{jk}^* &0  \\
        0 &C_{jk} &P_j &0  \\
        X_{jk} &0 &0 &0 
        \end{array}
        \right]
        + \mathcal{O}(\lambda^4)
\end{align}
obtained by tracing out the third detector in the system.

\begin{figure*}[t]
\centering
\includegraphics[width=13cm]{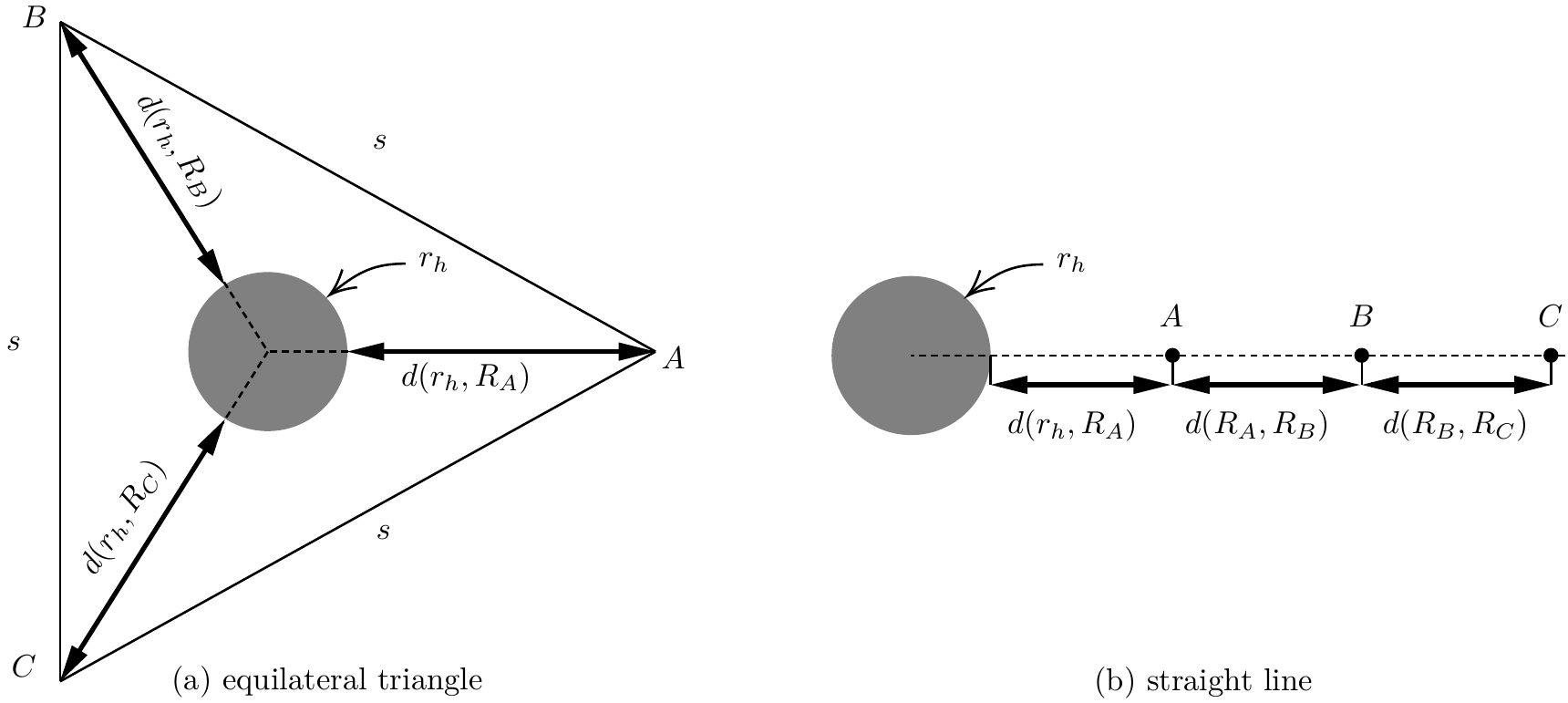}
\caption{Detector configurations we consider for tripartite entanglement harvesting.
(a) Equilateral triangle configuration. 
Detectors $A, B$, and $C$ are located at the corners of the equilateral triangle. 
(b) Straight line configuration. }
\label{fig:configurations}
\end{figure*}

\subsection{Entanglement measures}

The harvested bipartite and tripartite entanglement can be quantified by the \textit{negativity of entanglement} \cite{Vidal2002negativity} and the \textit{$\pi$-tangle} \cite{Ou2007}, respectively. 
Let $\rho_{jk}^{\top_j}$ be the partial transpose of density matrix $\rho_{jk}$ with respect to the $j$-component, where
\begin{align}
    \top_A\,:\, \ket{l_A, m_B} \bra{l'_A, m'_B} \mapsto \ket{l'_A, m_B} \bra{l_A, m'_B}
\end{align}
with similar definitions for $\top_B$ and $\top_C$.
Then the negativity of entanglement between detectors-$j$ and $k$ is defined by 
\begin{align}
    \mathcal{N}_{j(k)}
    &\coloneqq
        \dfrac{ ||\rho_{jk}^{\top_j}|| -1 }{2}\,,
\end{align}
where $||\cdot ||$ is the trace norm. 
Practically, this definition is equivalent to summing the magnitudes of all negative eigenvalues of the partial transposed density matrix $\rho_{jk}^{\top_j}$: 
\begin{align}
    \mathcal{N}_{j(k)}
    &=
        \sum_{\eta_i<0} |\eta_i|\,,
        \quad \eta_i \in \text{eigen}(\rho_{jk}^{\top_j})\,.
\end{align}
The negativity $\mathcal{N}_{j(k)}$ is nonzero if and only if the $2\times 2$ (or $2\times 3$) systems $j$ and $k$ are entangled. 

The $\pi$-tangle \footnote{We note that the $\pi$-tangle is a generalization of three-tangle \cite{Coffman:1999jd}, where concurrence is replaced by negativity. 
The advantage of using the $\pi$-tangle is that it can detect a genuine entanglement that the three-tangle might miss.} is defined by using the negativity: 
\begin{align}
    \pi
    \coloneqq
        \dfrac{ \pi_A + \pi_B + \pi_C }{3}\,,\label{eq:pitangle}
\end{align}
where 
\begin{subequations}
\begin{align}
    \pi_A 
    &=
        \mathcal{N}_{A(BC)}^2-\mathcal{N}_{A(B)}^2-\mathcal{N}_{A(C)}^2\,, \label{eqn:PiA} \\
    \pi_B 
    &=
        \mathcal{N}_{B(AC)}^2-\mathcal{N}_{B(A)}^2-\mathcal{N}_{B(C)}^2\,, \label{eqn:PiB} \\
    \pi_C 
    &=
        \mathcal{N}_{C(AB)}^2-\mathcal{N}_{C(B)}^2-\mathcal{N}_{C(A)}^2\,,\label{eqn:PiC}
\end{align}
\label{eq:SubPi}
\end{subequations}
and by definition, 
\begin{align}
    \mathcal{N}_{A(BC)}
    &=
        \dfrac{ ||\rho_{ABC}^{\top_A}|| -1 }{2}\,.
\end{align}
Note that $\mathcal{N}_{A(BC)}$ does not reflect the entanglement between detector-$A$ and the rest, $(BC)$, since it is $2\times 4$ and the PPT criterion is  sufficient but not necessary; 
it only provides the lower bound for entanglement. 
This means that $\mathcal{N}_{A(BC)}=0$ does not necessarily mean that entanglement between the subsystems is 0. 

Nevertheless, the $\pi$-tangle \eqref{eq:pitangle} quantifies the tripartite entanglement in the system when $\rho_{ABC}$ is a pure state. 
This is supported by a monogamy relation called the Coffman-Kundu-Wootters (CKW) inequality for negativity \cite{Coffman:1999jd, Ou2007}: 
\begin{align}
     \mathcal{N}^2_{A(B)} + \mathcal{N}^2_{A(C)} 
     \leq \mathcal{N}^2_{A(BC)}\,.
\end{align}
In general, however, this inequality will not hold when $\rho_{ABC}$ is mixed. 
In such  cases the $\pi$-tangle defined in \eqref{eq:pitangle} and \eqref{eq:SubPi} is only a lower bound for the tripartite entanglement. In order to obtain genuine tripartite entanglement for a mixed state, one needs to generalize the inequality to  \cite{Ou2007}
\begin{align}
    \mathcal{N}^2_{A(B)} + \mathcal{N}^2_{A(C)}
    \leq \textrm{min} \left[ \braket{ \mathcal{N}^2_{A(BC)} } \right]\,,
\end{align}
where $\braket{ \mathcal{N}^2_{A(BC)} } \coloneqq \sum_i p_i \mathcal{N}^2_{A(BC)}$ is the expectation value and the minimization is performed over all possible pure-state decompositions $\rho_{ABC}=\sum_i p_i \ket{\psi_i}\bra{\psi_i}$ \cite{Coffman:1999jd, Ou2007}. 
$\mathcal{N}^2_{A(BC)}$ for a given $\rho_{ABC}$ is the lower bound, $\mathcal{N}^2_{A(BC)} \leq \braket{ \mathcal{N}^2_{A(BC)} }$, for any pure-state decomposition since   negativity has the convexity property. 
Since such minimization is tedious, we will use the $\pi$-tangle \eqref{eq:pitangle} as a lower bound for mixed tripartite entanglement.

Since the parameter space for three different detectors is quite large, for simplicity  we will consider  them  to have identical energy gaps and couplings, which are defined in the reference frame of the detectors: $\Omega_j=\Omega$ and $\lambda_j\chi_j(\tau_j/\sigma)=\lambda\chi(\tau_j/\sigma)$ for all $j\in\{A,B,C\}$.  The detectors will be located at fixed distances outside of the horizon
\begin{equation}
    \sx_j(\tau_j) \coloneqq \left\{t=\frac{\tau_j}{\gamma_j},\ r=R_j,\ \varphi=\varphi_j\right\} \,,
\end{equation}
where $\gamma_j\coloneqq\sqrt{R_j^2-r_h^2}/\ell$ is the redshift factor of each detector.
The detectors will be arranged in one of the two configurations depicted in Fig.~\ref{fig:configurations}:  one in which the detectors are placed at the vertices of an a equilateral triangle; the other in which they are placed along a straight line.

The equilateral triangle configuration simplifies the form of the negativities and $\pi$-tangle when the detectors are identical \cite{DianaGaussianTripartite},  and so is a convenient configuration to consider due to its symmetry.
Since the three detectors have the same switching function $\chi(\tau/\sigma)$ and energy gap $\Omega$, the symmetry of the configuration allows us to write the elements of the density matrix \eqref{eq:rhoABC2ndOrder} as 
\begin{subequations}
    \begin{align}
        &C_{AB}=C_{BC}=C_{CA}\equiv C\,, \\
        &X_{AB}=X_{BC}=X_{CA}\equiv X\,, \\
        &P_A = P_B= P_C \equiv P 
    \end{align}
\end{subequations}
and so the negativities can be simply written as \cite{DianaGaussianTripartite}
\begin{align}
    &\mathcal{N}_{A(BC)}= \mathcal{N}_{B(CA)} = \mathcal{N}_{C(AB)} \notag \\
    &=
        \max 
        \left\{
            0,\,
            \dfrac{ \sqrt{ C^2 + 8|X|^2 } }{2} - \dfrac{C}{2} - P
        \right\} + \mathcal{O}(\lambda^4)\,, \\
    &\mathcal{N}_{A(B)} = \mathcal{N}_{A(C)} = \mathcal{N}_{B(C)} = \mathcal{N}_{B(A)} = \mathcal{N}_{C(A)} = \mathcal{N}_{C(B)} \notag \\
    &= 
        \max \{ 0,\, |X|-P \} + \mathcal{O}(\lambda^4) \,.\label{eq:equilateral bipartite negativity}
\end{align}
The $\pi$-tangle \eqref{eq:pitangle} is then 
\begin{align}
    \pi
    &=
        \max 
        \left\{
            0,\,
            \dfrac{ \sqrt{ C^2 + 8|X|^2 } }{2} - \dfrac{C}{2} - P
        \right\}^2 \notag \\
        &\quad - 2 \max \{ 0,\, |X|-P \}^2 + \mathcal{O}(\lambda^6) \,.\label{eq:equilateral triangle pitangle}
\end{align}
We easily see from \eqref{eq:equilateral bipartite negativity}  that   bipartite entanglement between detectors in the equilateral triangle configuration exists if the nonlocal correlation $|X|$ is greater than the transition probability $P$. 
In other words, if the transition probability is large enough due to, for example, intense Hawking radiation near an event horizon, then the detectors are unlikely to extract bipartite entanglement from the field. 
This is also true for the $\pi$-tangle -- if
the non-local correlations are insufficiently large, then the first term in
\eqref{eq:equilateral triangle pitangle} will vanish.

Note that, although a straight line configuration in Minkowski spacetime allows us to somewhat simplify the $\pi$-tangle in a similar manner \cite{DianaGaussianTripartite}, this is not the case in the BTZ spacetime since each detector arranged in a straight line in Fig.~\ref{fig:configurations}(b) experiences different redshift. 

In what follows, we will evaluate 
$\pi/\tilde\lambda^4$, in other words we shall compute the $\pi$-tangle per unitless coupling.

\begin{figure}[t]
	\centering
 \includegraphics[width=\linewidth]{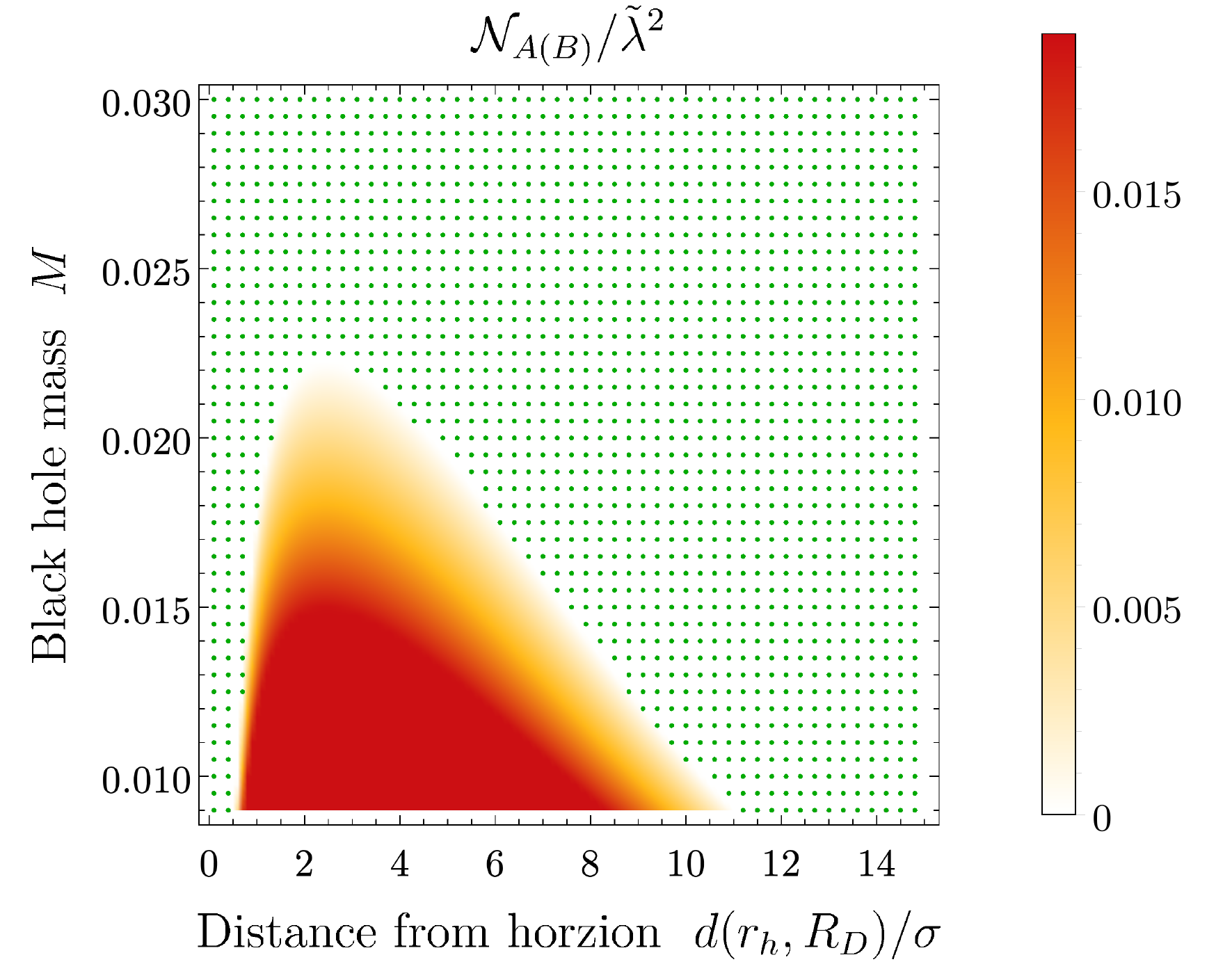}\\
 \vspace{1em}
	\includegraphics[width=\linewidth]{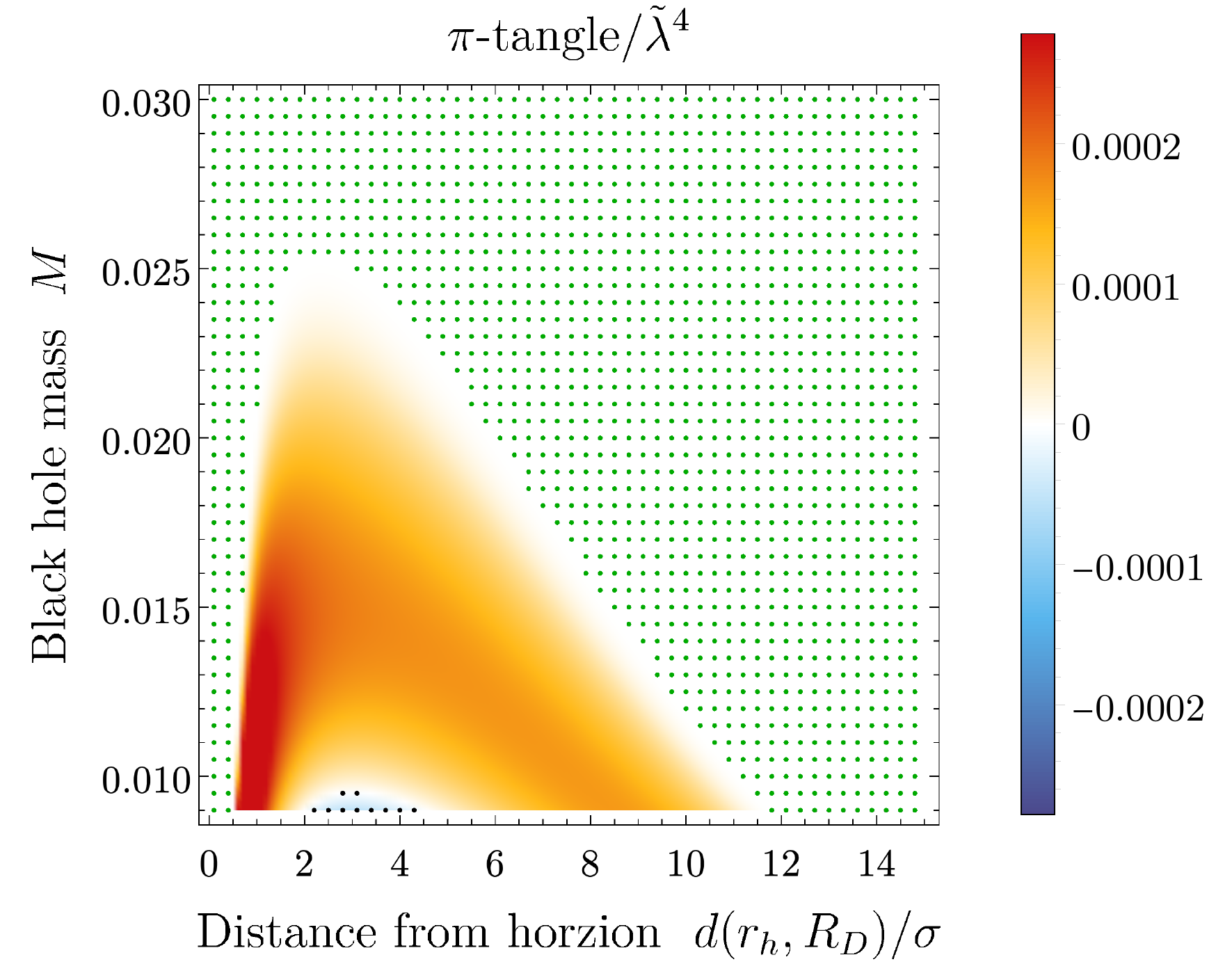}
 	\caption{%
		\textit{(Top:)} The bipartite negativity, $\mathcal{N}_{D(D')}$, between two of the detectors, and \textit{(bottom:)} the extracted $\pi$-tangle in the case of the equilateral triangle configuration depicted in Fig.~\ref{fig:configurations}(a) as a function of the mass of the black hole $M$ and the proper distance $d(r_h, R_D)/\sigma$ of the detectors from the horizon. 
        Here, the AdS length $\ell/\sigma=10$, the boundary condition $\zeta=1$, and the energy gap of the detectors $\Omega \sigma=1$. 
        In the top figure the green dots indicate the region where $\mathcal{N}_{A(B)}=0$ and in the bottom figure
        the green dots indicate the region where the $\pi$-tangle is zero. Black dots indicate the region where the $\pi$-tangle is negative.  
	}
	\label{fig:densityM}
\end{figure}

\section{Results}
\label{sec3}

We now explore the $\pi$-tangle for these two detector configurations to determine the conditions under which tripartite entanglement can be harvested. 
The switching function for the detectors is a Gaussian, which is expressed as $\chi(\tau_j/\sigma)=\exp(-\tau_j^2/2\sigma^2)$. 
See Appendix for the explicit formulae for $P_j, C_{jk}$, and $X_{jk}$. 
We shall vary the parameters of the configuration (detector spacing, energy gap) and the dimensionless black hole mass $M$ to search for regions of positive $\pi$-tangle.  
This latter condition ensures genuine entanglement between all three detectors, since the $\pi$-tangle only implies a lower bound on the magnitude of tripartite entanglement.
We shall see that tripartite harvesting  is possible well within the shadow regions  \cite{henderson2018harvesting} of bipartite entanglement. 
Moreover, we also find that some regions in parameter space have no tripartite entanglement shadow.



\subsection{Equilateral triangle configuration}

We first consider the equilateral triangle configuration shown in Fig.~\ref{fig:configurations}(a).  
To provide context, we first briefly present results for bipartite entanglement harvesting as there has been little exploration of the dependence entanglement harvested by a pair of detectors and the angle between them \cite{HendersonThesis}. 


In Fig.~\ref{fig:densityM}(top), we plot the bipartite negativity $\mathcal{N}_{D(D')}$ between two of the three detectors in the equilateral triangle configuration as a function of the black hole mass and the proper distance of the detectors from the horizon. 
We find that bipartite entanglement harvesting by detectors with an energy gap of $\Omega\sigma=1$ is only possible, at any distance from the horizon, if the mass of the black hole is smaller than $M\approx0.0225$. 
This is likely due to the fact that as the mass $M$ of the black hole increases, the proper separation $d(R_D, R_{D'})/\sigma$ between the two detectors also increases if one fixes the distance from the horizon, $d(r_h, R_D)/\sigma$. 
Since greater detector separations, $d(R_D, R_{D'})/\sigma$, decrease the correlations between the detectors, greater black hole masses will decrease the maximum distance from the horizon where entanglement harvesting is possible.

Moreover, the bipartite entanglement is zero when the detectors are very close to the horizon, which is caused by the high black hole temperature \cite{henderson2018harvesting, Kendra.BTZ}. 
The term $|X|$ in \eqref{eq:equilateral bipartite negativity} sharply decreases while the transition probability $P$ increases with temperature. 
Hence, the entanglement shadow still exists even when detectors are located at a different angle $\varphi_j$.


However, we see from Fig.~\ref{fig:densityM}(bottom),  where we now consider the $\pi$-tangle between all three detectors, 
that harvesting tripartite entanglement is indeed possible for black hole masses larger than those that allow for bipartite entanglement harvesting. 
The fact that the detectors can extract tripartite entanglement when the bipartite one is vanishing can be seen from \eqref{eq:equilateral triangle pitangle}. 
We also find it  possible to extract tripartite entanglement at larger distances from the horizon than is possible in the bipartite case, which corresponds to greater detector separations, in agreement with  previous results in flat spacetime \cite{DianaGaussianTripartite}. 
We note that for this particular value of energy gap, $\Omega\sigma=1$, the $\pi$-tangle becomes zero for all detector distances from the horizon when the black hole mass is larger than $\approx0.0255$. 
Furthermore, we find that when the black hole mass is less than $\approx0.01$, the $\pi$-tangle becomes negative for moderate detector distances, despite being positive for both smaller and larger distances. 
Since the $\pi$-tangle \eqref{eq:pitangle} puts a lower bound on the tripartite entanglement, we are unable to gain any information about multivariate correlations between the three detectors in this region 
(where $\pi\leq 0$) 
of the parameter space, and instead focus our attention on  regions where the $\pi$-tangle is positive. 
For this reason we do not explore black holes with smaller masses.

\begin{figure}[t]
	\centering
 \includegraphics[width=\linewidth]{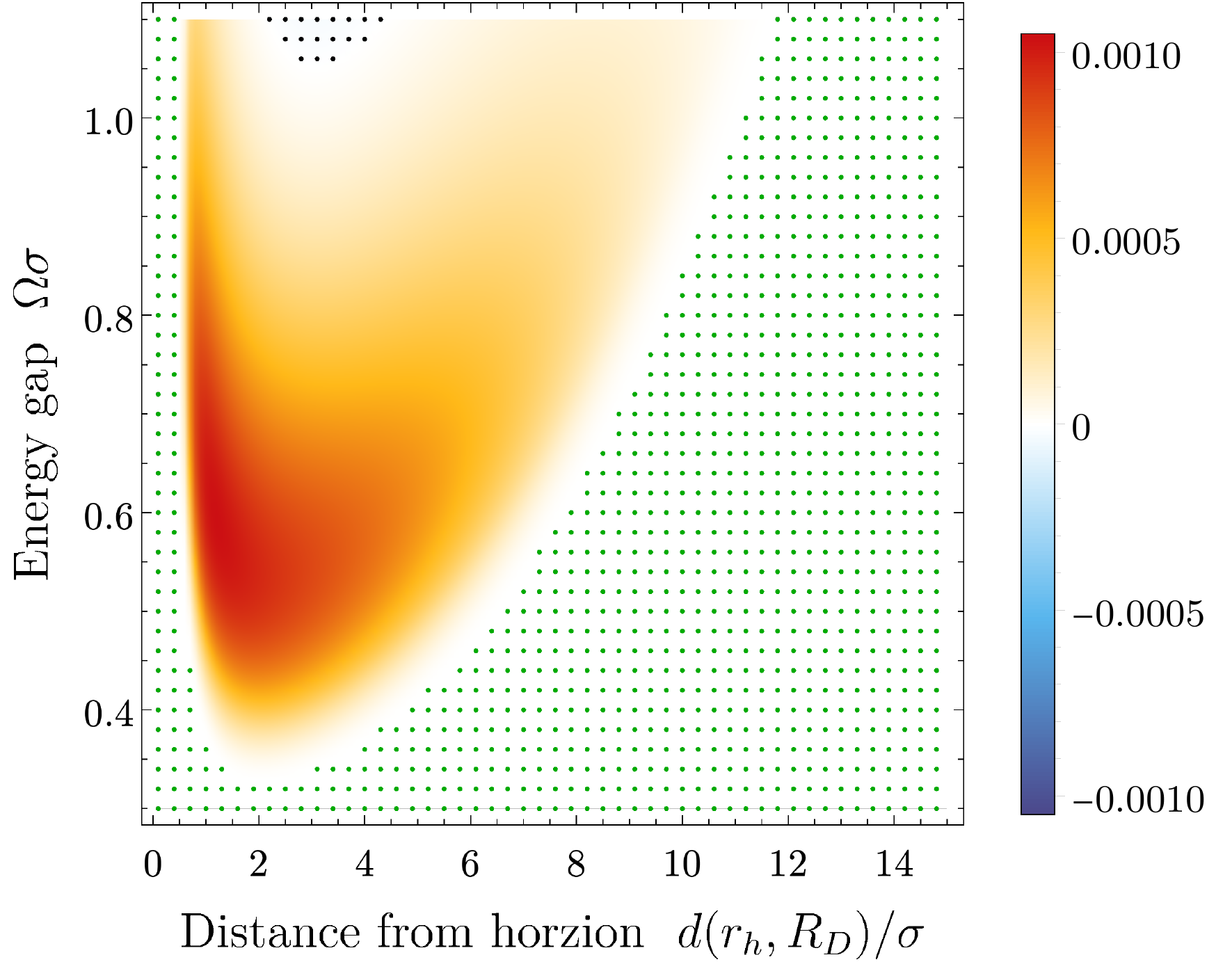}
 	\caption{%
		In the case of three detectors placed in an equilateral triangle configuration outside of a BTZ black hole, the $\pi$-tangle is calculated as a function of the energy gap $\Omega \sigma$ of the detectors and the proper distance of the detectors from the horizon, $d(r_h, R_D)/\sigma$. 
        The mass of the black hole is $M=0.01$. 
        The green dots indicate the region where the $\pi$-tangle is zero and the black dots indicate the region where the $\pi$-tangle is negative.
	}
	\label{fig:densityOmega}
\end{figure}

Turning next to exploring the dependence of the $\pi$-tangle on the other parameters of the problem, the AdS length $\ell$ and the detector energy gap $\Omega$, we find that tripartite entanglement harvesting is only guaranteed (via a positive $\pi$-tangle) in a small region of the parameter space. 
 A plot of the $\pi$-tangle for varying AdS length $\ell$ and equilateral triangle size yields a graph very similar to that of the lower panel in Fig.~\ref{fig:densityM}: for particular AdS lengths within a sensitive range $\ell/\sigma \leq 16$ 
tripartite entanglement can be harvested across approximately the same range of proper distances from the black hole horizon. 
Plotting in Fig.~\ref{fig:densityOmega} the $\pi$-tangle for varying energy gap $\Omega$ and distance, we see that tripartite entanglement can be harvested across a broad range of distances from the horizon
for $0.4 \leq \Omega\sigma \leq 1$. 

These plots show that entanglement slowly decreases as the size of the triangle increases, with the $\pi$-tangle remaining nonzero even when the detectors are spacelike separated.




\begin{figure}[t]
	\centering
	\includegraphics[width=\linewidth]{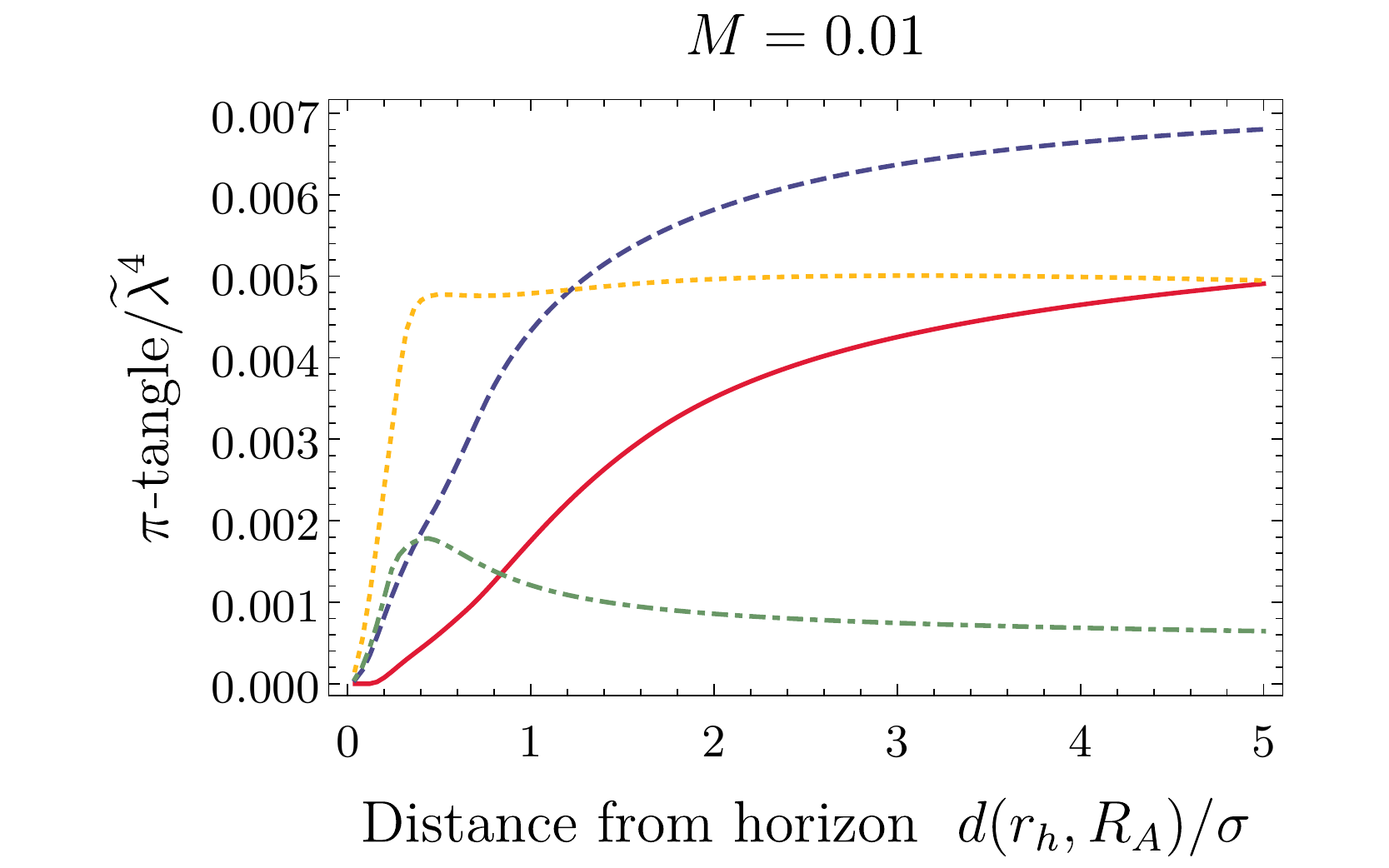}\\
    \vspace{1em}
    \includegraphics[width=\linewidth]{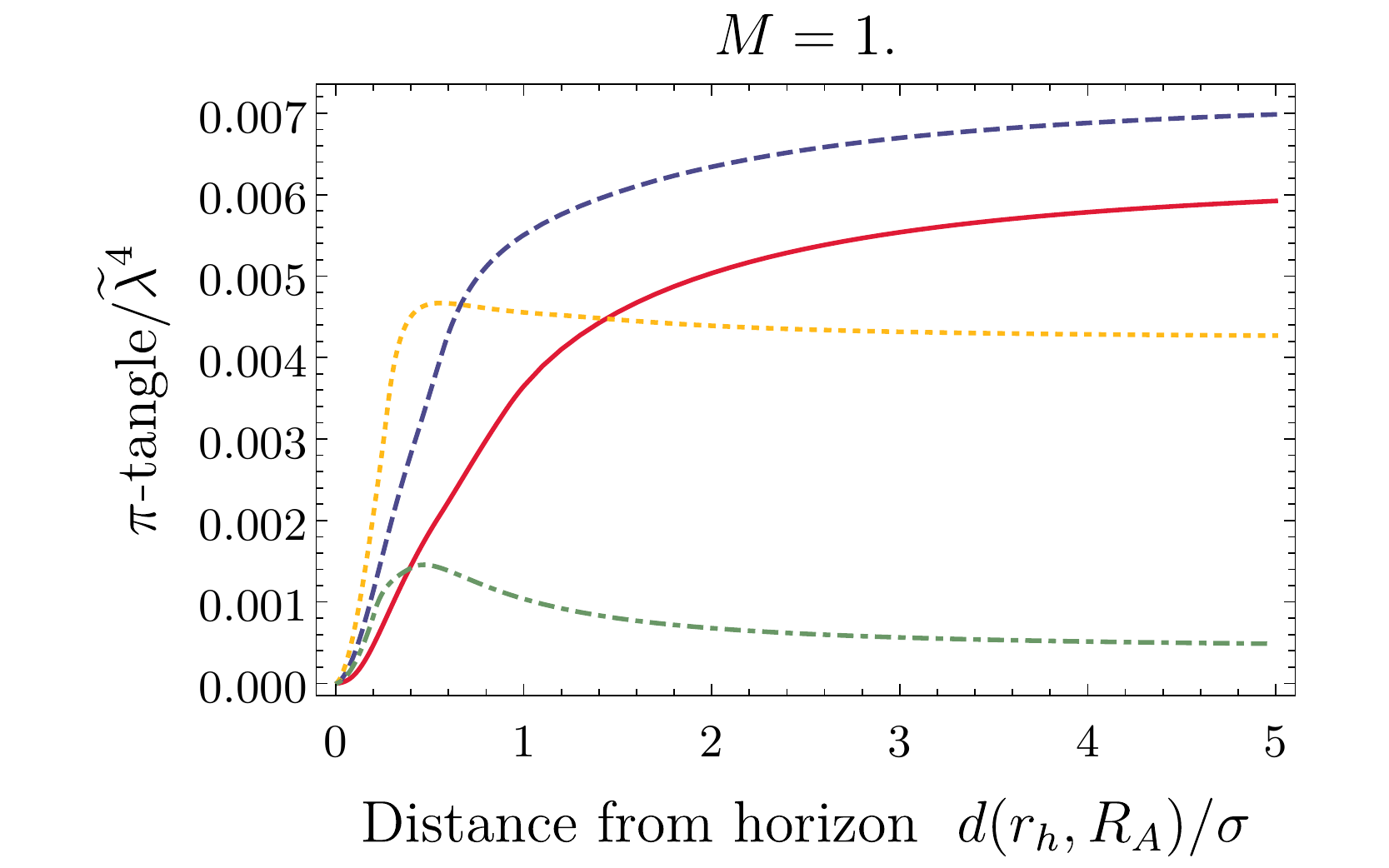}\\
    \includegraphics[width=\linewidth]{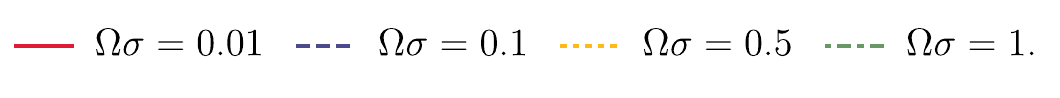}
 	\caption{%
		In the case of three detectors placed along a straight line outside of a BTZ black hole as shown in Fig.~\ref{fig:configurations}(b), the $\pi$-tangle is calculated as a function of the proper distance of the closest detector $A$ to the horizon, $d(r_h, R_A)/\sigma$ for a black hole mass of \textit{(top)}: $M=0.01$ and \textit{(bottom)}: $M=1$. 
        Here, the proper distance between the detectors remains constant at $d(R_A, R_B)/\sigma= d(R_B, R_C)/\sigma=1$. 
        The AdS length $\ell/\sigma=10$ and the boundary condition $\zeta=1$. 
	}
	\label{fig:line_pi}
\end{figure}

\begin{figure}[ht]
	\centering
	\includegraphics[width=\linewidth]{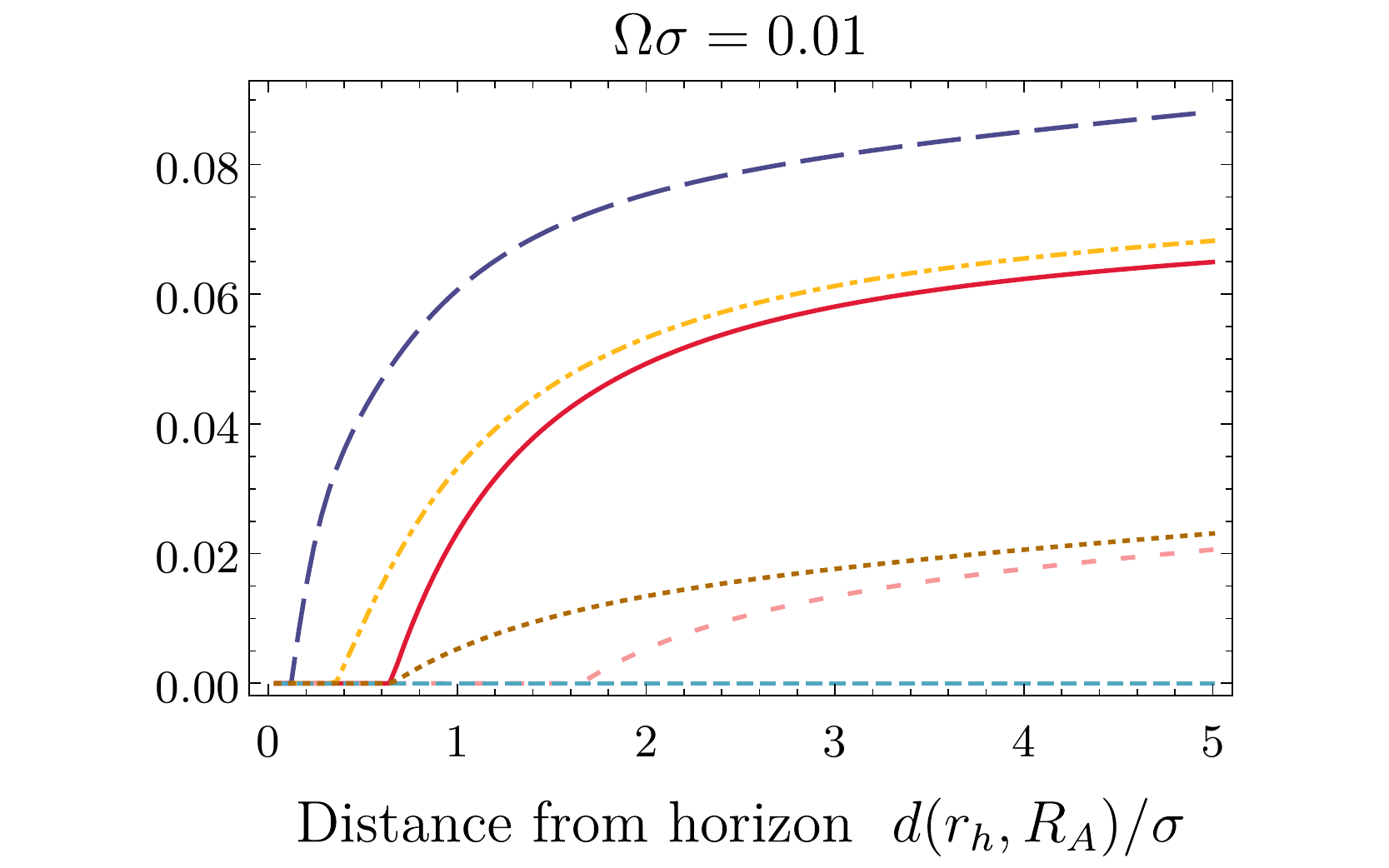}\\
    \vspace{1em}
    \includegraphics[width=\linewidth]{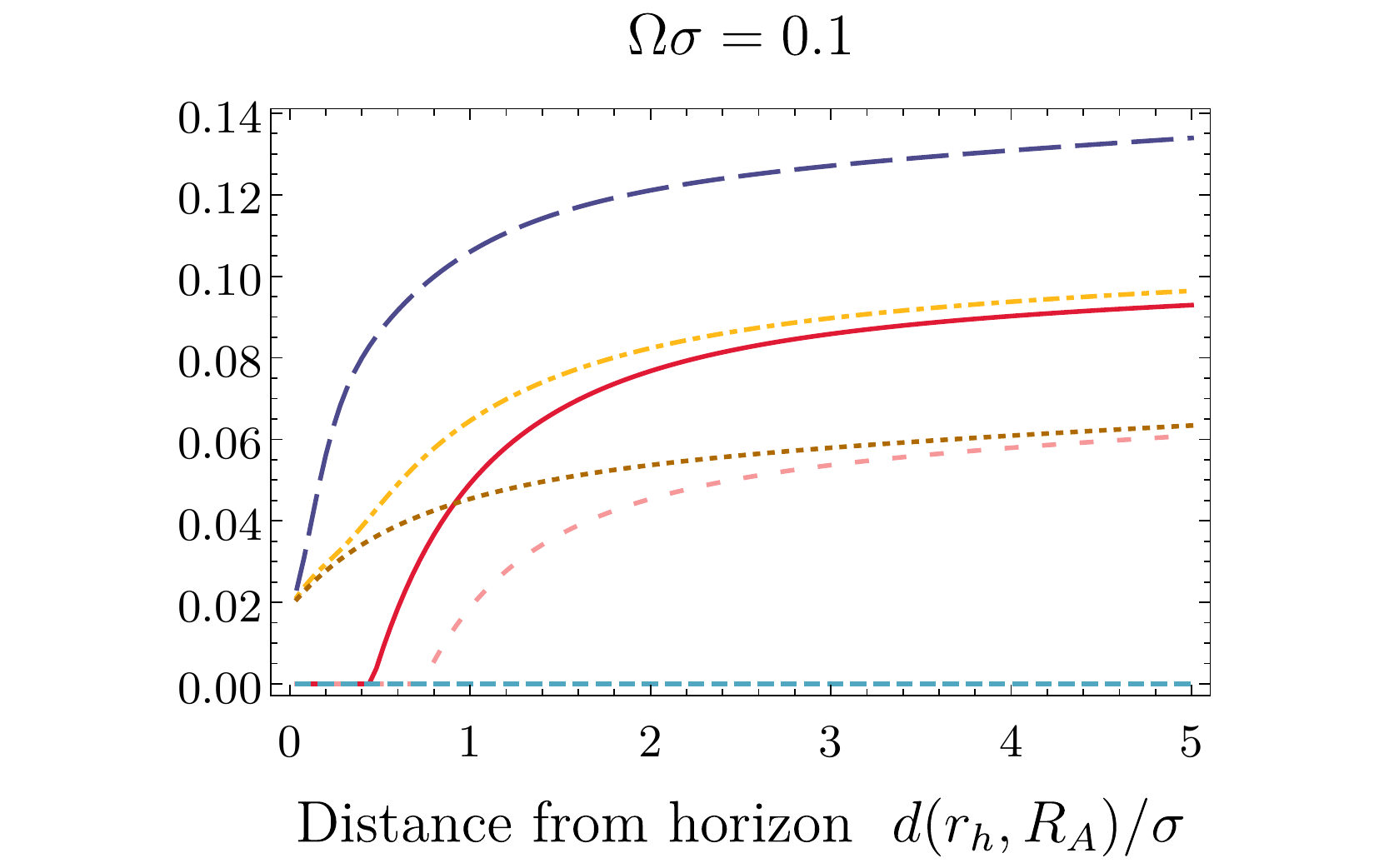}
    \includegraphics[width=0.8\linewidth]{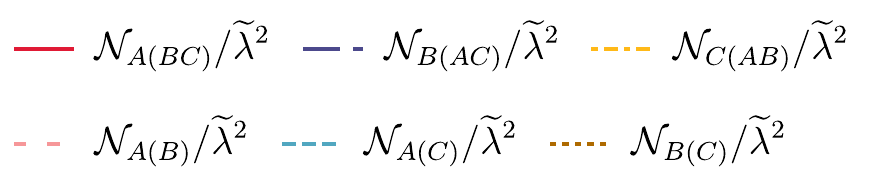}
 	\caption{%
		In the case of three detectors placed along a straight line outside of a BTZ black hole [Fig.~\ref{fig:configurations}(b)], the three tripartite and three bipartite negativities are calculated as a function of the proper distance of the closest detector $A$ to the horizon, $d(r_h, R_A)/\sigma$ for a detector energy gap of \textit{Top:} $\Omega\sigma=0.01$ and \textit{Bottom:} $\Omega\sigma=0.1$. 
        Here, the proper distance between the detectors remains constant at $d(R_A, R_B)/\sigma=d(R_B, R_C)/\sigma=1$. 
        The AdS length $\ell/\sigma=10$, the boundary condition $\zeta=1$, and the mass of the black hole $M=0.01$. 
	}
	\label{fig:line_neg}
\end{figure}

\subsection{Straight line configuration}

By placing the three detectors along a line intersecting the center of the black hole [Fig.~\ref{fig:configurations}(b)], we gain further insight into the distinctions between the  bipartite and tripartite cases.
In  the bipartite case  the entanglement shadow was manifest when one of the detectors was placed close to the horizon \cite{henderson2018harvesting}. 
However,  we find that the situation is very different in the tripartite case. 

In Fig.~\ref{fig:line_pi}, we fix the proper distance between the detectors to be $d(R_A, R_B)/\sigma=d(R_B,R_C)/\sigma= 1$ and plot the $\pi$-tangle as a function of the proper distance from the black hole horizon to the nearest detector for two different black hole masses and differing energy gaps. 
We find that, provided the energy gap of the detectors is large enough, tripartite entanglement can be harvested at distances arbitrarily close to the black hole horizon.
 In striking distinction to the bipartite case, tripartite  entanglement often has no entanglement shadow in the line configuration, and,  unlike in the case of the equilateral triangle configuration, it is quite easy to find regions in   parameter space where this occurs.  
The minimum value of the detector energy gap guaranteeing near horizon tripartite entanglement harvesting depends on the mass of the black hole. 
Specifically, we find that for  $M=0.01$ and  $\Omega\sigma=0.01$ the $\pi$-tangle becomes zero when the first detector is very close at $d(r_h,R_A)/\sigma \lessapprox 0.08$. 
It remains positive at these very close distances when the mass of the black hole is large, $M=1$, shown in the lower panel of Fig.~\ref{fig:line_pi}.  
We also note that the $\pi$-tangle asymptotes to a constant value, which depends on $\Omega$ and $M$, in the limit $d(r_h,R_A)/\sigma \to \infty$, as expected since the BTZ spacetime is asymptotically $\ads{3}$. 


 We now explicitly show that the $\pi$-tangle is positive where the bipartite entanglement is zero. 
Choosing two cases  $(M, \Omega \sigma)=(0.01, 0.01)$ and $(0.01, 0.1)$ in Fig.~\ref{fig:line_pi}(top), we plot their bipartite and tripartite negativities in \eqref{eq:SubPi} as a function of the proper distance of detector $A$ from the horizon in Fig.~\ref{fig:line_neg}. 
 For both values of the energy gap $\Omega \sigma$ considered, we find that each $\pi_j\,(j\in \{ A, B, C \})$ in \eqref{eq:pitangle} is positive even when some of the bipartite negativities $\mathcal{N}_{D(D')}$ vanish. 
Consider $\pi_A$ \eqref{eqn:PiA} as an example. 
We see that $\mathcal{N}_{A(BC)}$ remains positive while $\mathcal{N}_{A(B)}=\mathcal{N}_{A(C)}=0$, which suggests that $\pi_A>0$ in the bipartite entanglement shadow. 
This is also true for $\pi_B$ and $\pi_C$ if they possess the bipartite entanglement shadow, and therefore the $\pi$-tangle is positive.\footnote{Since the tripartite negativity is a lower bound, $\mathcal{N}_{A(BC)}=0$ does not mean that there is no entanglement between detector $A$ and the $(BC)$ subsystem; there is very likely to be some, since detector $B$ is more entangled with $(AC)$ and detector $C$ is more entangled with $(AB)$ then detector $B$ is entangled with detector $C$ alone.} 
We point out that $\mathcal{N}_{B(C)}$ does not vanish in Fig.~\ref{fig:line_pi}(bottom) since detectors $B$ and $C$ are still far away from the horizon when $d(r_h, R_A)/\sigma=0$. 
In fact, this is one of the main differences between the equilateral triangle and the straight line configurations. 
Due to its symmetry, the $\pi$-tangle for the equilateral triangle configuration is characterized by only three elements, namely $P$, $C$, and $X$.
However, this is not the case for the straight line configuration, resulting in the detector furthest from the event horizon being least affected by the black hole. 
Consequently, this leads to the existence of a nonvanishing $\pi$-tangle near the horizon as shown in Fig.~\ref{fig:line_pi}. 
Nevertheless, for a sufficiently large $\Omega \sigma$, tripartite entanglement can be extracted in the bipartite entanglement shadow, as in the case of equilateral triangle configuration, which suggests that the harvested entanglement is of the GHZ type.

\begin{figure}[t]
	\centering
	\includegraphics[width=\linewidth]{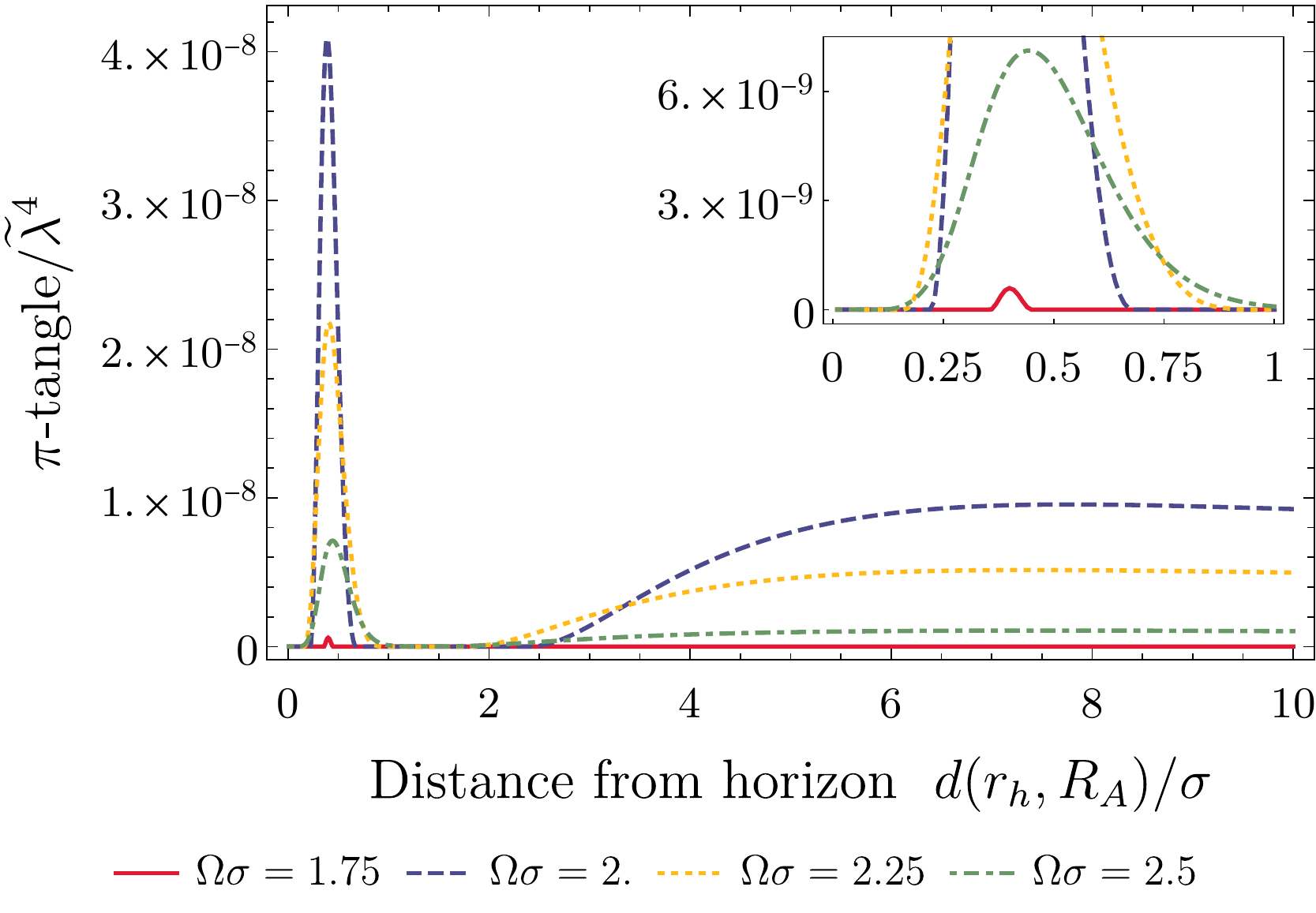}
 	\caption{%
		In the case of three detectors placed along a straight line outside of a BTZ black hole [Fig.~\ref{fig:configurations}(b)], the $\pi$-tangle is calculated as a function of the proper distance of the closest detector $A$ to the horizon, $d(r_h, R_A)/\sigma$ for various values of the energy gap of the detectors. 
        Here, the proper distance between the detectors remains constant at $d(R_A, R_B)/\sigma=d(R_B, R_C)/\sigma=5$. 
        The AdS length $\ell/\sigma=10$, the boundary condition $\zeta=1$, and the mass of the black hole $M=0.01$.  The inset shows details of the near horizon spike in $\pi$-tangle for $\Omega\sigma=1.75$ and $\Omega\sigma=2$. 
    }
	\label{fig:line_largespacing_pi}
\end{figure}

We note from Fig.~\ref{fig:line_pi} that a local maximum exists in $\pi/\tilde{\lambda}^2$  for some $\Omega \sigma$. 
In fact, for some choices of parameters, this maximum leads to an interesting result. 

In Fig.~\ref{fig:line_largespacing_pi} we again plot the $\pi$-tangle  of the detectors in the straight line configuration as a function of the proper distance of the nearest detector from the horizon, but now for larger detector separation $d(R_A, R_B)/\sigma=d(R_B, R_C)/\sigma=5$. 
Unlike  the previous case, as the detectors move towards the horizon, the $\pi$-tangle becomes zero followed by a sharp spike before it drops to zero again. 
These spikes in the $\pi$-tangle are larger in magnitude than its asymptotic value at $d(r_h, R_A)/\sigma \to \infty$. 
  
\begin{figure}[t]
	\centering
	\includegraphics[width=\linewidth]{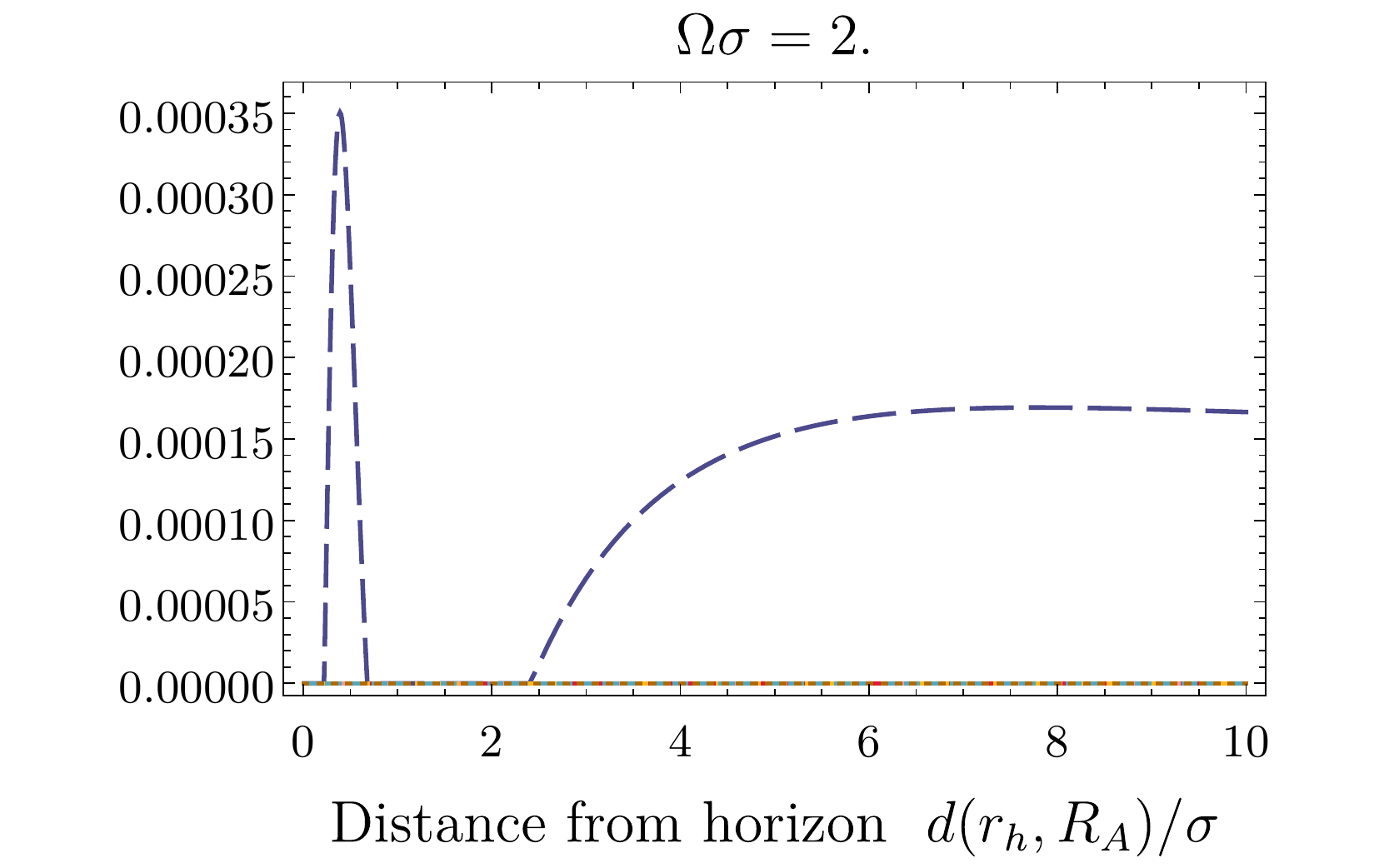}\\
    \vspace{1em}
    \includegraphics[width=\linewidth]{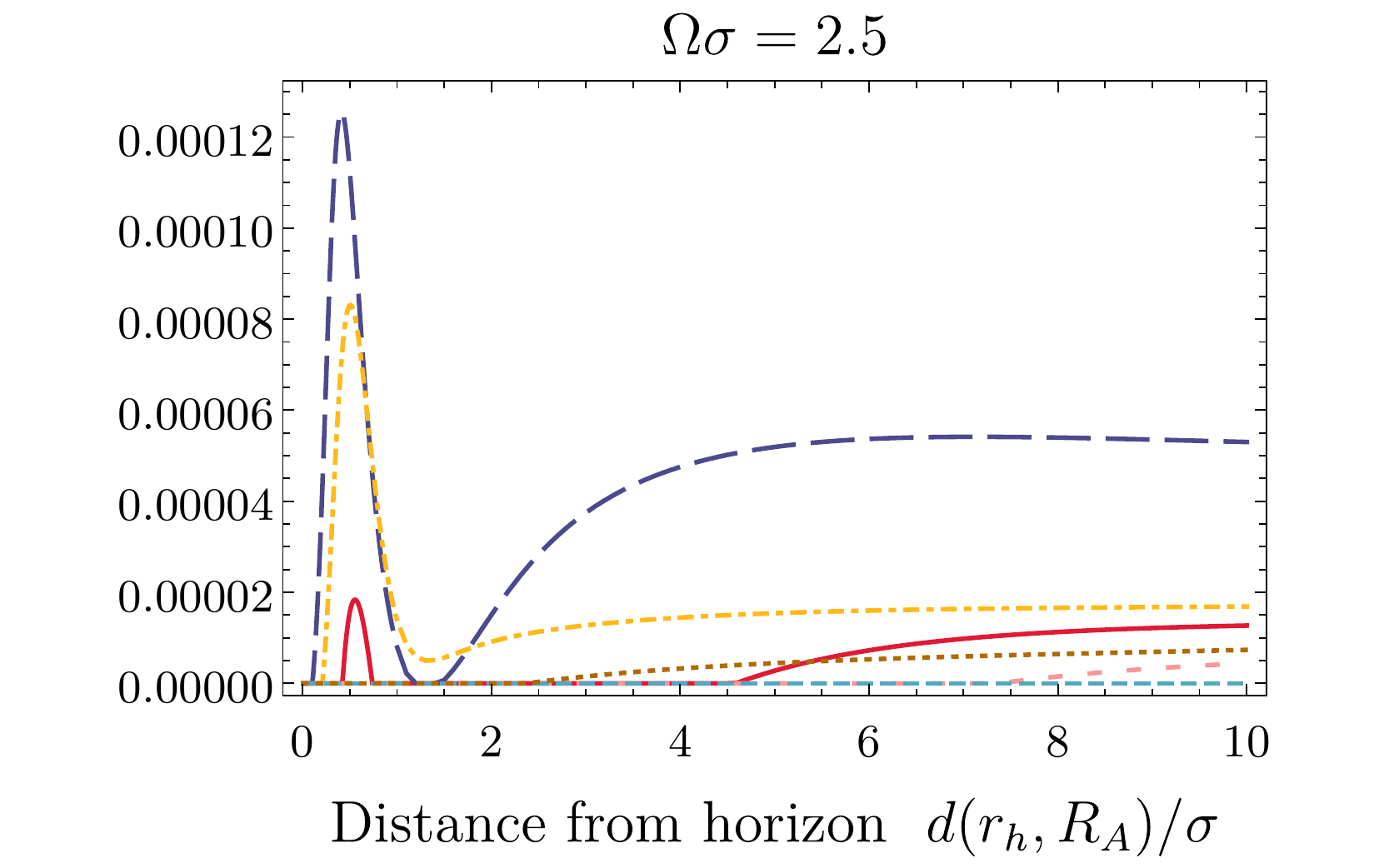}
    \includegraphics[width=0.8\linewidth]{Laura_Figures/Legend_Negs.pdf}
 	\caption{%
		In the case of three detectors placed along a straight line outside of a BTZ black hole [Fig.~\ref{fig:configurations}(b)], the three tripartite and three bipartite negativities are plotted  as a function of the proper distance of the closest detector $A$ to the horizon, $d(r_h, R_A)/\sigma$ for a detector energy gap of \textit{Top:} $\Omega\sigma=2$ and \textit{Bottom:} $\Omega\sigma=2.5$.
        Here, the proper distance between the detectors remains constant at $d(R_A, R_B)/\sigma=d(R_B, R_C)/\sigma=5$. 
        The AdS length $\ell/\sigma=10$, the boundary condition $\zeta=1$, and the mass of the black hole $M=0.01$.
	}
	\label{fig:line_largespacing_neg}
\end{figure}

 To better understand the origin and properties of the near-horizon spike in the $\pi$-tangle, we plot the bipartite and tripartite negativities in Fig.~\ref{fig:line_largespacing_neg}. 
We first note that the large spike in the $\pi$-tangle occurs at values of $d(r_h,R_A)$ where the bipartite negativities are zero, i.e.~inside the bipartite entanglement shadow of all three detectors, meaning that the tripartite entanglement extracted from the field in this region of the spacetime is of the GHZ type. 
Consequently, the $\pi$-tangle results from the tripartite negativities, and when the energy gap of the detectors is small enough, only from $\mathcal{N}_{B(AC)}$, which puts a lower bound on the entanglement between detector $B$ and the $(AC)$ subsystem. 
As the energy gap of the detector is increased, the remaining two tripartite negativities also contribute to near-horizon $\pi$-tangle.

\begin{figure*}
    \centering
    \includegraphics[width=0.32\linewidth]{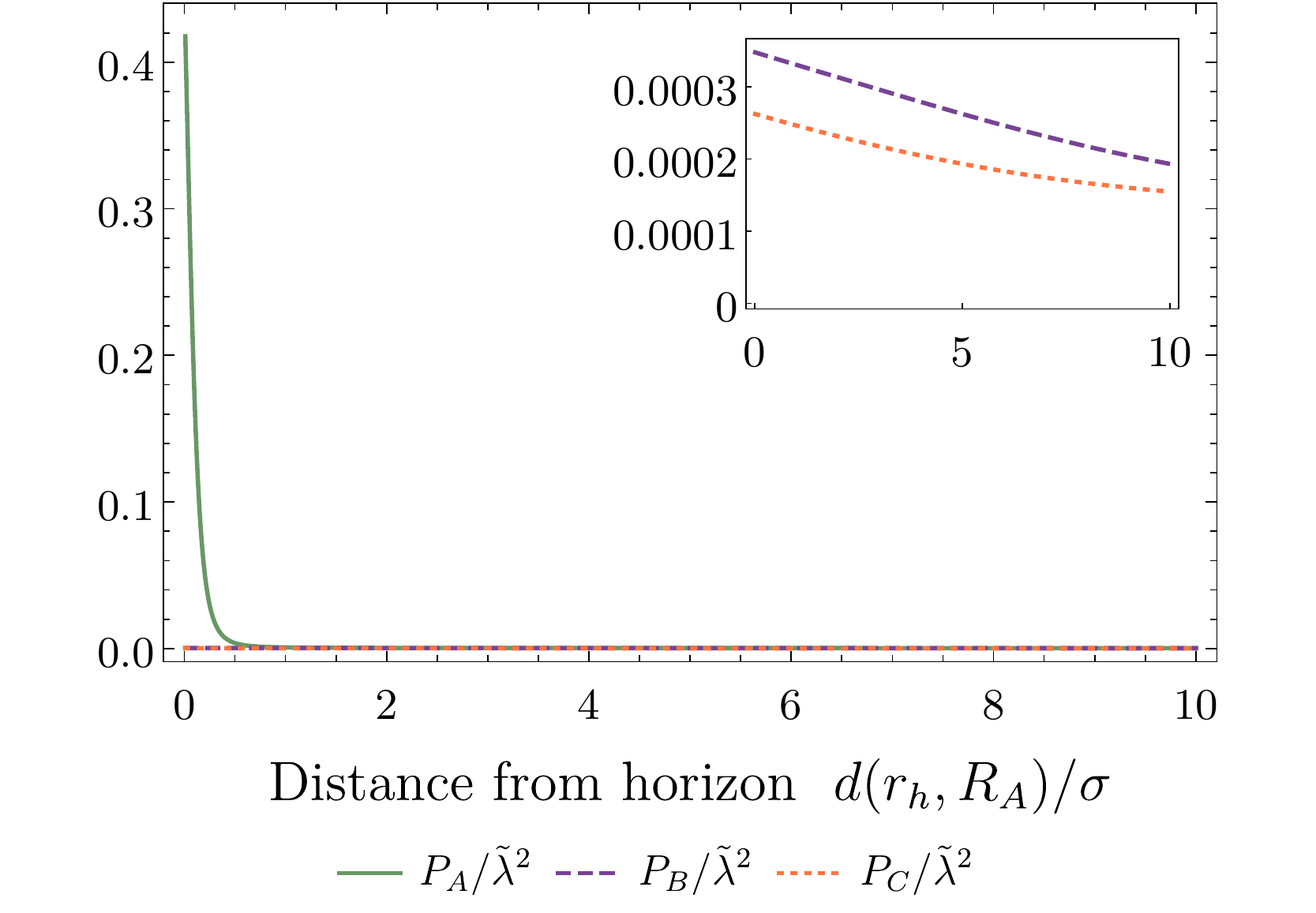}
    \includegraphics[width=0.32\linewidth]{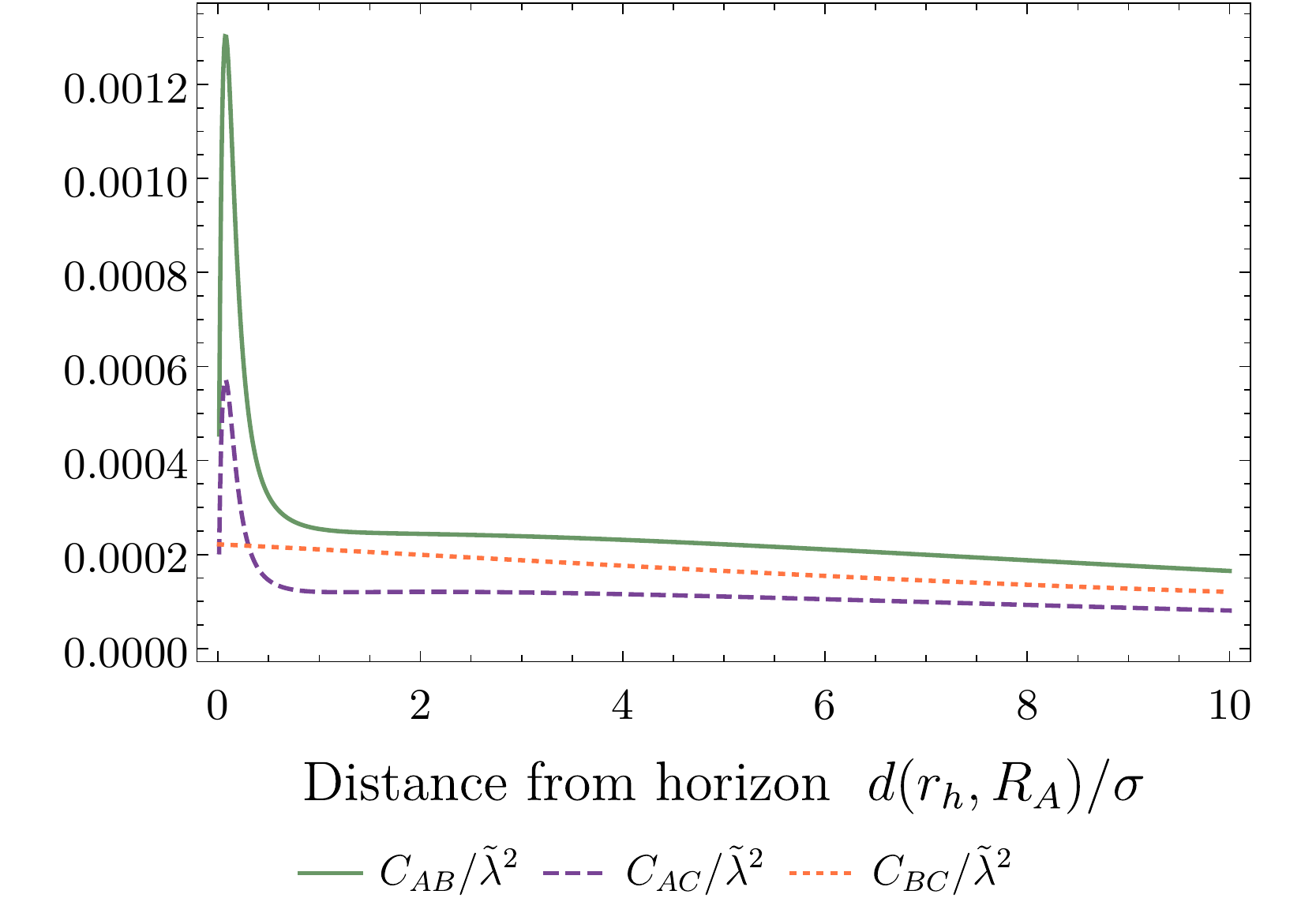}
    \includegraphics[width=0.32\linewidth]{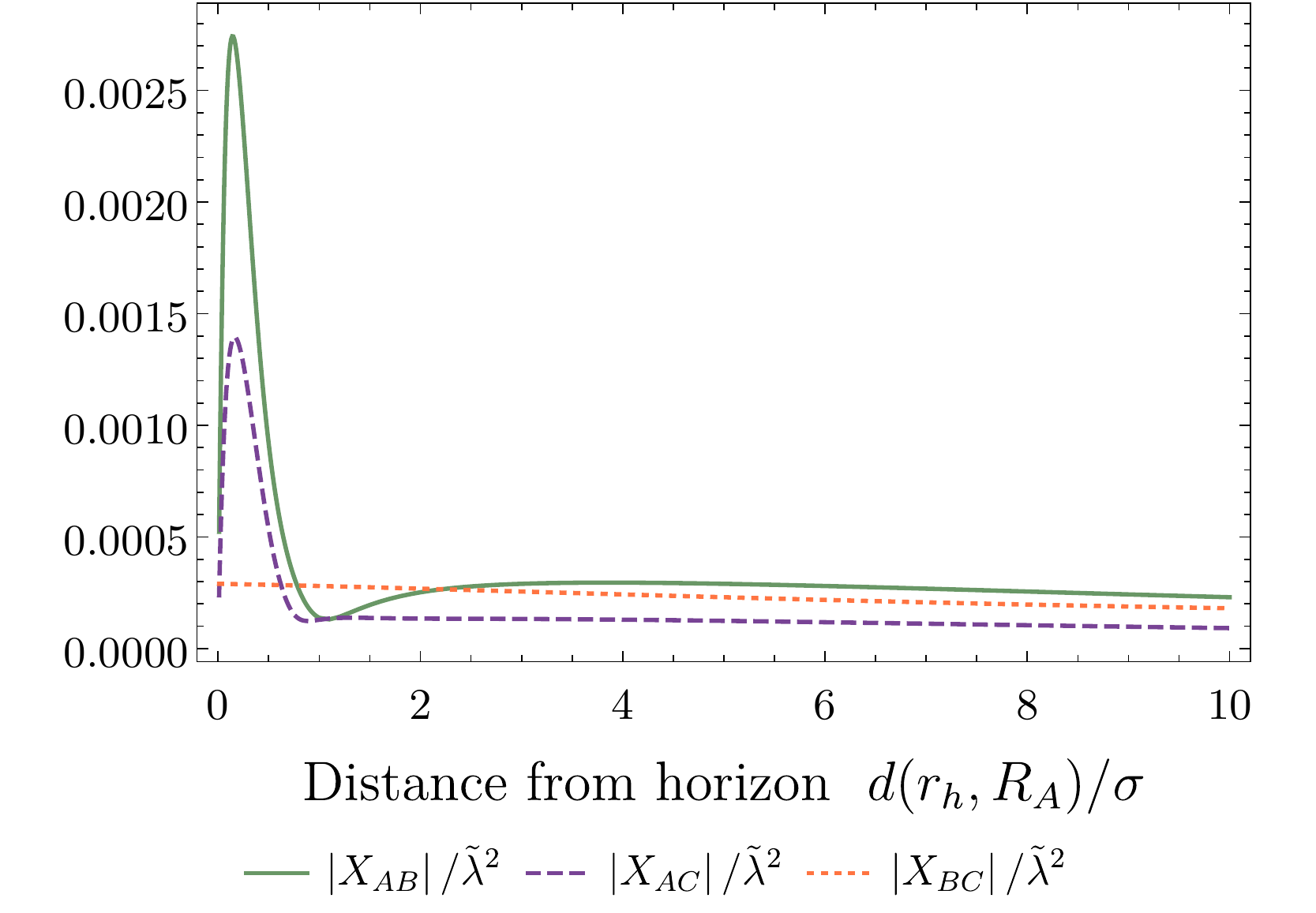}
    \caption{In the case of three detectors placed along a straight line outside of a BTZ black hole [Fig.~\ref{fig:configurations}(b)] with $\ell/\sigma=10$, and $M=0.01$, the matrix elements of Eq.~\eqref{eq:rhoABC2ndOrder} are calculated as a function of $d(r_h, R_A)/\sigma$, for \textit{Left:} the transitions probabilities $P_D/\tilde{\lambda}^2$, \textit{Center:} the pairwise detector correlations $C_{D D'}/\tilde{\lambda}^2$, and \textit{Right:} the pairwise nonlocal correlations $|X_{DD'}|/\tilde{\lambda}^2$. 
    Here, the proper distance between the detectors remains constant at $d(R_A, R_B)/\sigma=d(R_B, R_C)/\sigma=5$ and the energy gap of the detectors is $\Omega\sigma=2.5$. 
    The inset in the left figure shows details of $P_B$ and $P_C$. 
    }
    \label{fig:MatixElements}
\end{figure*}

The bipartite and tripartite negativities are functions of the matrix elements of Eq.~\eqref{eq:rhoABC2ndOrder}, which we plot in Fig.~\ref{fig:MatixElements} for the same detector configuration as in Fig.~\ref{fig:line_largespacing_pi}. 
We find that as detector $A$ approaches the horizon, both the detector correlations $C_{DD'}$ and the nonlocal correlations $|X_{DD'}|$ between detectors $A$ and $B$ and between $A$ and $C$ also show a near-horizon spike. 
Here, the spike in the detector correlations $C_{DD'}$ is smaller and occurs closer to the horizon than that of the nonlocal correlations $|X_{DD'}|$. 
The correlations between detectors $B$ and $C$ do not exhibit spikes, as they are still relativity far from the horizon, $d(r_h,R_B)/\sigma=d(r_h,R_A)/\sigma+5$. 
This spike in pairwise correlations is directly responsible for the spike in the tripartite negativities shown in Fig.~\ref{fig:line_largespacing_neg}. 
Such a spike is not seen in the bipartite negativities, since the transition probability of detector $A$, $P_A/\tilde{\lambda}^2$, is very large near the horizon, due to the local field temperature, which will prevent any entanglement between detector $A$ and $B$ (or $C$). 
Since detectors $B$ and $C$ remain far from the horizon, their transition probabilities remain small, and as a result, the three detectors can have tripartite entanglement. 
We also note that a near horizon spike does not appear in the pairwise correlations when $\Omega\sigma<1$.


\section{Conclusion}

Our key finding is that black holes affect distinct  properties of the quantum vacuum in quite distinct ways. 
In particular, tripartite scalar field vacuum correlations have notably different properties from their bipartite counterparts. 
Specifically, by investigating the tripartite entanglement harvesting protocol from a trio of UDW detectors outside of a BTZ black hole, we found that the detectors are able to harvest tripartite entanglement in larger regions of the parameter space than the bipartite case.

When the three detectors are arranged in an equilateral triangle, we find that tripartite entanglement can be harvested at larger distances from the horizon and for larger masses than is possible in the bipartite case. 
We also find that  the bipartite negativity between any two of the detectors goes to zero at a larger proper distance than is the case for the $\pi$-tangle.
This is also seen for the mass of the black hole:  if the detectors are at a fixed proper distance from the horizon, and the mass of the black hole is increased, the bipartite negativity goes to zero before $\pi$-tangle does. 
However, when the mass of the black hole is sufficiently small, the $\pi$-tangle becomes negative at moderate distances from the horizon, and we can no longer guarantee tripartite harvesting, even though bipartite harvesting is possible. 
We find a similar result  for small values of the AdS length and large detector energy gaps, indicating that for this configuration, it is difficult to find regions of the parameter space where tripartite entanglement harvesting is guaranteed, in particular when the curvature is very high.

A negative $\pi$-tangle does not indicate a failure of the entanglement harvesting protocol, but rather occurs as a  consequence of using it on a mixed state \cite{DianaGaussianTripartite}. 
This is a reminder that in the case of a mixed state, as in the case of the bipartite entanglement harvesting protocol, the $\pi$-tangle is a lower bound for the tripartite entanglement of the system. 
In other words, in regions in   parameter space where the $\pi$-tangle is non-positive, no conclusions can be drawn about the resultant tripartite entanglement between the detectors. 
It is only in regions where the $\pi$-tangle is positive that we can conclude the detectors can harvest tripartite entanglement.

Previous studies have shown the BTZ black hole does not have a mutual information shadow, and that a pair of UDW detectors can harvest mutual information when they are arbitrarily close to the horizon\cite{Kendra.BTZ}. 
This raises the question about extractable quantum discord from the black hole vacuum, particularly because there is a strong link between bipartite discord and tripartite entanglement\cite{BROWN2013153}. 
It would be interesting to see how this quantity is affected by the presence of a horizon.


If the three detectors are arranged on a straight line of constant $\varphi$ in the BTZ coordinates,  the detectors can harvest tripartite entanglement even when the bipartite entanglement cannot be extracted. 
Furthermore, detectors can still harvest tripartite entanglement even when the detector nearest to the black hole is extremely close to the horizon. 
This is similar to the mutual information harvesting scenario \cite{Kendra.BTZ} in which an extremely high black hole temperature prevents two detectors from extracting total correlations from the field.

We also found a tripartite entanglement peak near a horizon in the straight line configuration. 
This is a distinctive feature of tripartite entanglement since bipartite entanglement does not contribute to this peak. 
It originates from an enhancement of pairwise correlations between detectors when one of them is close to the horizon, but is not detected in the bipartite entanglement since the local noise near the horizon is very large. 
However, because $\pi$-tangle ``averages out'' the local noise of three detectors rather than two, the boost in correlations is able to dominate over the local noise, resulting in a positive $\pi$-tangle.
We also note that this near-horizon peak occurs when the detectors have large proper separations and will have minimal communication over the duration of the interaction, meaning that this tripartite entanglement must come from the vacuum state of the field.

We note that the results we have presented are all for Dirichlet boundary conditions ($\zeta = 1$). 
While it would be of some interest to fully explicate the other boundary conditions ($\zeta=0$ and $\zeta=-1$), we have found they yield qualitatively similar results for the triangular configuration, and so expect no new features to emerge in more general settings.

A more interesting   avenue for future research would be  to classify the harvested tripartite entanglement. 
In general, tripartite entanglement can be classified into four categories: GHZ, W, biseparable, and fully separable states \cite{DurThreeQubits, Acin.Mix.Tripartite}. 
With $\pi$-tangle, one can show the existence of the GHZ-type entanglement since the bipartite negativity between qubits is an entanglement monotone. 
However, a complete characterization of tripartite entanglement for a mixed state is not straightforward. 
We leave an investigation of this problem for future study.

\section*{Acknowledgments}

This work was supported in part by the Natural Sciences and Engineering Research Council of Canada, by the Asian Office of Aerospace Research and Development Grant No. FA2386-19-1-4077, and by the ARC Centre of Excellence for Engineered Quantum Systems (EQUS) CE170100009.

\appendix
\section{$P_j, C_{jk}$, and $X_{jk}$ in BTZ spacetime}
The elements of the density matrix \eqref{eq:elements general} for the UDW detectors with a Gaussian switching function in BTZ spacetime are known to be \cite{henderson2018harvesting, Kendra.BTZ}
\begin{align}
    P_j
    &=
        \dfrac{\lambda^2 \sigma^2}{ 2 }
        \int_{\mathbb{R}} \dd x\,
        \dfrac{ e^{ -\sigma^2 (x-\Omega)^2 } }{ e^{ x/T_j } +1 } \notag \\
        &-\zeta \dfrac{ \lambda^2 \sigma }{2 \sqrt{2 \pi} }
        \text{Re}
        \int_0^\infty \dd x\,
       \dfrac{ e^{-a_j x^2} e^{-\ii \beta_j x} }{ \sqrt{ \cosh \alpha_{j,0}^+ - \cosh x } }
       \notag \\
        &+\dfrac{ \lambda^2 \sigma }{ \sqrt{2 \pi} }
        \sum_{n=1}^\infty 
        \text{Re}
        \int_0^\infty \dd x\,
        e^{ -a_j x^2 } e^{ -\ii \beta_j x } \notag \\
        &\times 
        \kako{
            \dfrac{1}{ \sqrt{ \cosh \alpha_{j,n}^- -\cosh x } }
            -
            \dfrac{\zeta}{ \sqrt{ \cosh \alpha_{j,n}^+ -\cosh x } }
        } , \label{eq:BTZ transition prob}
\end{align}
where $T_j=r_{h}/2 \pi \ell^2 \gamma_j$ is the local temperature at $r=R_j$ and 
\begin{subequations}
    \begin{align}
        a_j
        &\coloneqq 
            \dfrac{ \ell^4 \gamma_j^2 }{ 4\sigma^2 r_{h}^2 },~
            \beta_j \coloneqq 
            \dfrac{ \ell^2 \gamma_j \Omega }{ r_{h} }\,, \\
        \alpha_{j,n}^\pm
        &\coloneqq
            \text{arccosh}
            \kagikako{
                \dfrac{r_{h}^2}{ \gamma_j^2 \ell^2 }
                \kako{
                    \dfrac{r^2}{ r_{h}^2 } \cosh 
                    \kagikako{
                        \dfrac{r_{h}}{\ell} 2\pi n
                    } \pm 1
                }
            } ,
    \end{align}
\end{subequations}
as well as 
\begin{align}
    &C_{jk}
    =
        K_- \sum_{n=-\infty}^\infty 
        \text{Re}\int_0^\infty \dd x\,
        e^{ -a_{jk} x^2 }
        e^{ -\ii \beta_+ x }  \notag \\
        &\times 
        \kako{
            \dfrac{ 1 }{ \sqrt{ \cosh \alpha_{jk,n}^- - \cosh x } }
            -
            \dfrac{ \zeta }{ \sqrt{ \cosh \alpha_{jk,n}^+ - \cosh x } }
        }, \label{eq:LabAppendix} \\
    &X_{jk}
    =
        - K_+ 
        \sum_{n=-\infty }^\infty 
        \int_0^\infty \dd x\,
        e^{ -a_{jk} x^2 } \cos (\beta_- x) \notag \\
        &\times 
        \kako{
            \dfrac{ 1 }{ \sqrt{ \cosh \alpha_{jk,n}^- - \cosh x } }
            -
            \dfrac{ \zeta }{ \sqrt{ \cosh \alpha_{jk,n}^+ - \cosh x } }
        },
\end{align}
where 
\begin{subequations}
\begin{align}
    &a_{jk}
    \coloneqq
        \dfrac{ \gamma_j^2 \gamma_k^2 }{ 2\sigma^2 (\gamma_j^2 + \gamma_k^2) } 
        \dfrac{ \ell^4 }{ r_{h}^2 }, \\
    &\beta_\pm
    \coloneqq
        \dfrac{ \gamma_j \gamma_k (\gamma_j \pm \gamma_k) }{ \gamma_j^2 + \gamma_k^2 }
        \dfrac{ \ell^2 }{r_{h}}
        \Omega\,, \\
    &K_\pm
    \coloneqq
        \dfrac{ \lambda^2 \sigma \sqrt{ \gamma_j \gamma_k } }{ 2 \sqrt{\pi} \sqrt{ \gamma_j^2 + \gamma_k^2 } }
        \exp
        \kagikako{
            -\dfrac{ \Omega^2 \sigma^2 (\gamma_j \pm \gamma_k)^2 }{ 2 (\gamma_j^2 + \gamma_k^2) }
        }, \\
    &\alpha_{jk,n}^\pm
    \coloneqq \notag \\
        &\text{arccosh}
        \kako{
            \dfrac{r_h^2}{ \gamma_j \gamma_k \ell^2 }
            \kagikako{
                \dfrac{R_j R_k}{ r_h^2} 
                \cosh 
                \kako{
                    \dfrac{r_h}{\ell} (\Delta \varphi + 2\pi n)
                }
                \pm 1
            }
        }.
\end{align}

\end{subequations}

\bibliography{ref}

\end{document}